\documentclass[aps,pre,twocolumn,nofootinbib]{revtex4-1}
\usepackage{amsmath,bm,epsfig}
 \usepackage[dvips]{color}

\def\Fbox#1{\vskip1ex\hbox to 8.5cm{\hfil\fboxsep0.3cm\fbox{%
  \parbox{8.0cm}{#1}}\hfil}\vskip1ex\noindent}  

\definecolor{g-blue}{rgb}{0.83,0.95,1}
\definecolor{g-yellow}{rgb}{1,1,0.7}
\definecolor{g-green}{rgb}{0.9,1,0.9}
\definecolor{green}{rgb}{0,0.6,0}
\definecolor{cyan}{rgb}{0,0.7,0.7}
\definecolor{black}{rgb}{0,0,0}
\definecolor{grey}{rgb}{0.4 ,0.4 ,0.4 }

\begin{document}

\title{Analytical modeling for the heat transfer in sheared flows of
  nanofluids}

\author{Claudio Ferrari$^1$, Badr Kaoui$^2$, Victor S. L'vov$^3$,
  Itamar Procaccia$^3$, Oleksii Rudenko$^{2,4}$, J.H.M. ten Thije
  Boonkkamp$^4$, and Federico Toschi$^{2,4,5}$} \affiliation{
  $^1\,$ISIS R\&D Srl, Viale G. Marconi 893, Roma
  I-00146, Italy.\\
  $^2\,$Department of Applied Physics, Eindhoven University of Technology, Eindhoven, 5600MB, The Netherlands.\\
  $^3\,$Department of Chemical Physics, The Weizmann Institute of Science, Rehovot 76100, Israel.\\
  $^4\,$Department of Mathematics \& Computer Science, Eindhoven University of Technology, Eindhoven, 5600MB, The Netherlands.\\
  $^5\,$CNR-IAC, Via dei Taurini 19, 00185 Rome, Italy.}

 \begin{abstract}
   We developed a model for the enhancement of the heat flux by spherical and
   elongated nano-particles in sheared laminar flows of nano-fluids.
   Besides the heat flux carried by the nanoparticles the model accounts for the
   contribution of their rotation to the heat flux inside and outside the
   particles. The rotation of the nanoparticles has a twofold
   effect, it induces a fluid advection around the particle and it
   strongly influences the statistical distribution of particle
   orientations. These dynamical effects, which were not included in existing thermal
   models, are responsible for changing the thermal properties of flowing fluids as compared
to quiescent fluids. The proposed model is strongly
   supported by extensive numerical simulations, demonstrating a
   potential increase of the heat flux far beyond the
   Maxwell-Garnet limit for the spherical nanoparticles. The road ahead
   which should lead towards robust predictive models of heat flux enhancement is
   discussed.
 \end{abstract}
\maketitle



\section{Introduction}
The enhancement of heat transfer by embedded nano-particles in a fluid
subjected to a temperature gradient is an important issue that is
expected to help in technological application like solar heating, and
various cooling devices, including miniaturized computer
processors. Accordingly, many experiments were conducted in the last
couple of decades \cite{Rev2011a,Rev2011b,Rev2010a,Rev2009a,Rev2009b},
with a breakthrough announcement from a group at Argonne National
Laboratory, who studied water and oil-based nanofluids containing
copper oxide nanopaticles, and found an amazing 60\% enhancement in
thermal conductivity for only a 5\% volume fraction of nanoparticles
\cite{96ECLTL}. Subsequent research has however generated what
Ref. \cite{06KK} referred to as ``an astonishing spectrum of
results''. Results in the literature show sometime enhancement in the
thermal conductivity compared to the prediction of the Maxwell-Garnett
effective medium theory, and sometime values that are less than the
same prediction
\cite{Rev2011a,Rev2011b,Rev2010a,Rev2009a,Rev2009b}. Confusingly
enough, these discrepancies occur even for the same fluid and the same
size and composition of the nano-particles.

Some of these conflicting results were explained by either the
formation of percolated clusters of particles (for the case of
enhancement with respect to the Maxwell-Garret prediction) or by
surface (Kapitza) resistance (for the case of reduction with respect
to the same prediction) \cite{Rev2010a}. Interestingly enough,
enhancement is typically seen in {\em quiescent} fluids, and one
expects that the agglomeration of clusters will become impossible in
flows, be them laminar or turbulent. Of course, in technological
application flowing nano-fluids may be the rule rather than the
exception. Thus our aim in this article is to develop models of heat
flux in flowing nano-fluids, where the flow can be either laminar or
turbulent. In such systems we cannot expect an enhancement of heat
flux due to the agglomeration of particles (at least at low volume
fractions), and for the sake of simplicity we will assume that there
is no Kapitza resistance. Since it was reported in the literature that
elongated nano-particles are superior to spheres in quiescent
nanofluids \cite{97Nan}, we will study both spheres and
spheroids~\footnote{A spheroid, or ellipsoid of revolution is a
  quadric surface obtained by rotating an ellipse about one of its
  principal axes; in other words, an ellipsoid with two equal
  semi-diameters.} in flowing nanofluids. We will argue that
generically elongation may not be advantageous at all, and will
explain why.

For completeness, we will begin by studying quiescent nanofluids under
a temperature gradient. We will present extensive numerical simulation
that will demonstrate a very good agreement with the Maxwell-Garnett
theory for spherical nano-particles and with the generalization of Nan
et al \cite{97Nan} for elongated particles. In the case of flowing
nanofluid we will offer a model that will provide expressions for the
heat transfer for different values of the aspect ratio of the
particles, for different volume fractions and for laminar and
turbulent flows.

The structure of the paper is as follows.  In the next
Section~(\ref{s:model}) we formulate the problem, describe the equations
of motion and the
numerical procedure, and discuss the dilute suspension
approximation.  In Section~\ref{s:3Body} we describe the results of
numerical simulations of spherical and spheroidal particles in a fluid
at rest and in a shear flow.  In Section~\ref{s:4Model} we develop an
analytical model of heat transfer characteristic, i.e. the Nusselt
number, and analyze the model predictions of heat flux enhancement in
two limiting cases of very strong Brownian diffusion and a weak one.

\section{\label{s:model} Nanofluids in the dilute limit}
In this section we formulate the model for dilute nanofluids
laden with elongated spheroidal nanoparticles. This includes the basic
equation of motion for the velocity and temperature fields and the boundary
conditions (with constant velocity and temperature gradients far away
from particles). We also present details of the numerical simulations and their
validation.

\subsection{\label{ss:PF}Formulation of the problem and the flow geometry}
\begin{figure}[!t]
  \begin{center}
    \includegraphics[width=8.5cm]{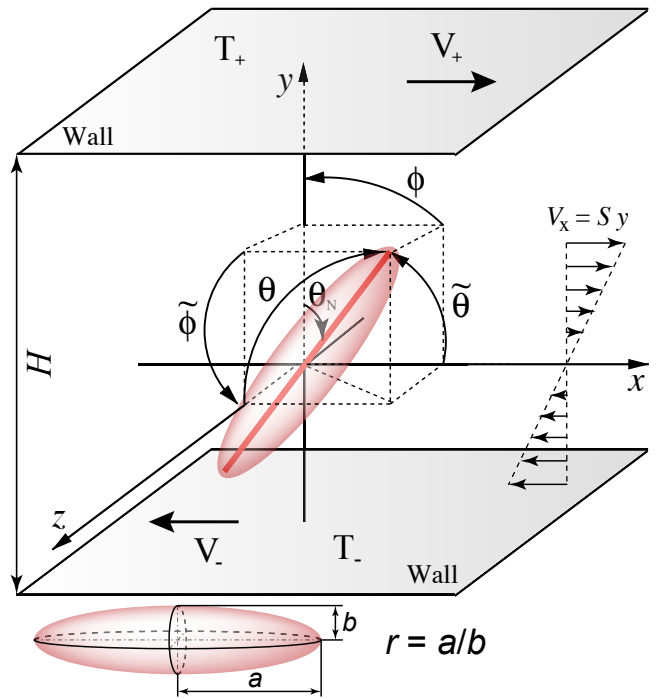}
  \end{center}
  \caption{\label{geometry} The particle in a plane Couette flow. The
    largest particle axis is shown as a (red) rod.  The velocity field
    (far from the particle) is $(Sy,0,0)$ where the shear $S = (V_{+} -
    V_{-})/H$ in the particle-free flow with $H$ being the distance
    between the walls. For the theoretical development we assume the
    creeping flow conditions as in \cite{Jef, Leal1971, Hinch1972}.}
\end{figure}

When the nanofluid is very dilute one can disregard the effect of one
particle on the other and consider the nanofluid as an ensemble of
noninteracting particles. One of these is shown in
Fig. \ref{geometry}. The center of the spheroidal particle is
at the position $x=y=z=0$ in the middle of a plane Couette flow
between two $H$-separated horizontal parallel walls that move in the
$\hat x$-direction with opposite velocities $V_{\pm}=\pm V/2$.  Due to
the symmetry, the forces are balanced, and the particle neither
migrates nor collides with the walls; the particle's center remains at
$(0,0,0)$. This allows us to eliminate in the numerics any effect of the particle
sweeping parallel to the walls.  Note that this effect is anyway absent in homogeneous cases.

We employ periodic boundary conditions along the $x$ and the $z$ direction. This results in
a periodic replication of the computational box (together
with the particle) in the $XZ$-plane, leading to a ``monolayer'' of an
infinite number of periodically distributed particles in the
$XZ$-plane. Remarkably enough (and by reasons that will be explained
below), this allows us to reproduce in numerics with a \emph{single
  particle} effects of \emph{a finite volume fraction} $\varphi$ at least
up to $\varphi \lesssim 20\div 30\%$.

The walls are kept at fixed temperatures $T_\pm = T \pm \Delta /2$.
In the absence of particles these boundary conditions give rise to a
vertical temperature gradient $\nabla_y T=\Delta/H$ and shear
$S=S_{xy}=V/H$ (see Fig.~\ref{geometry}).

The spheroidal particle's semi-axes are $a$ (the longest) and $b$ (the
shortest), $a \geq b$ (Fig.~\ref{geometry}). The thermal diffusivity
inside the particle is $\chi_{\rm p}$. The carrier fluid has a kinematic
viscosity $\nu$ and a thermal diffusivity $\chi_{\rm f}$.  For
simplicity, at this stage we neglect the effect of gravity and the
particle mass.  The co-ordinate system and polar angles are given by:
\begin{subequations}
  \label{eq:AnglesDefs}
  \begin{eqnarray}
    n_x =\,& \cos\widetilde\theta &\,= \sin{\theta} \sin{\phi}\,, \\
    n_y =\,& \sin\widetilde\theta \sin\widetilde\phi &\,= \sin{\theta} \cos{\phi} = \cos\theta_{\textrm N}\,, \\
    n_z =\,& \sin\widetilde\theta \cos\widetilde\phi  &\,= \cos{\theta}\,,
  \end{eqnarray}
\end{subequations}
where $n_i$ is the projection of the unit-vector $\widehat{\bm n}$
onto the axis $i$, $i = \{x, y, z\}$, and $\theta_{\textrm N}$ is the
angle between the particle's largest axis and the $y$-axis.

In the absence of particles, changing the intensity of the laminar
shear flow does not change the heat flux since there is no velocity
orthogonal to the wall and the system is homogeneous in the direction
parallel to the wall. In the presence of a particle, the shear induces
it to rotate, the simplest case being a spherical particle which
rotates at a constant angular velocity $\Omega_z = S/2$ (provided the
particle radius is much smaller than the distance to the wall).  This
rotation induces a vertical flow motion near the particle.  The
modification of the velocity profile has three consequences. The first
directly affects the heat convection in the fluid through a convective
contribution $v_y T$. The second comes from the particle rotation
which brings up the hotter side of the particle during its
rotation. The third changes the heat conduction due to a modification
of the temperature profiles at the surface of the particle.

In the case of non spherical particles the presence of a shear has
also a strong influence of the statistical distribution of particle
orientations.

\subsection{Dimensionless parameters}

To fix the notation we list here the most important dimensionless
parameters involved in the problem:
\begin{subequations}
  \label{DimNumDefs}
  \begin{eqnarray}
    \label{def:r}
    \mbox{aspect ratio:}\quad\quad\ r &\equiv& a/b\,,  \quad\ r \ge 1\,,\\
    \label{def:k}
    \mbox{diffusivities ratio:}\quad\quad\ k &\equiv& \chi_{\rm p}/\chi_{\rm f}\,, \\
    \label{def:Pr}
    \mbox{Prandtl number:}\quad\ \ \mathrm{Pr} &\equiv& \nu/\chi_{\rm f}\,, \\
    \label{def:Rep}
    \mbox{Reynolds number:}\quad \mathrm{Re} &\equiv& (2a)^2 S / \nu\,, \\
    \label{def:Pep}
    \mbox{Peclet number:}\quad \mathrm{Pe} &\equiv& (2a)^2 S / \chi_{\rm p}\ .
  \end{eqnarray}
\end{subequations}
Note that the particle's largest axis defines our length-scale.
Also,
\begin{equation}
  \label{Pe2Re}
  \mathrm{Pe}= \mathrm{Re}\,\mathrm{Pr} / k\ .
\end{equation}

\subsection{\label{ss:ME}Basic equations of motion}
The dynamical effects previously mentioned can be quantitatively
studied by solving the following system of equations for the
temperature in the fluid, $T_{\rm f}$, and in the particle, $T_{\rm
  p}$,
\begin{subequations}
  \label{basicEqs}
  \begin{eqnarray}
    \label{Eq1}
    \frac{\partial T_{\rm f}}{\partial t}+ (\bm u \cdot \bm \nabla) T_{\rm f} &=& \chi_{\rm f} \nabla^2 T_{\rm f}\,,
    \quad
    \frac{\partial T_{\rm p}}{\partial t} = \chi_{\rm p} \nabla^2 T_{\rm p} \ .~~~~
  \end{eqnarray}
  together with the boundary conditions,
  \begin{eqnarray}
    \label{Eq1bc}
    \mbox{at the surface:~~~}  & T_{\rm f} = {T}_{\rm p}\,,  &     \chi_{\rm f} {\nabla} T_{\rm f} =   \chi_{\rm p} {\nabla} {T}_{\rm f}\,, \nonumber\\
    \mbox{on the walls:~~~~~}  &  T_{\rm f}(x,0,z) = T_{-}\,,  &  T_{\rm f}(x,H,z) = T_{+}\ . \nonumber
\end{eqnarray}

The fluid velocity in the whole domain can be found by solving the
incompressible Navier-Stokes equations:
\begin{equation} \label{Eq2} \frac{\partial \bm u_{\rm f}}{\partial t}
  + (\bm u_{\rm f} \cdot \bm \nabla) \bm u_{\rm f} =
  -\frac{1}{\rho_{\rm f}}\bm \nabla p + \nu \nabla^2 \bm u_{\rm f}\,,
  \quad \bm \nabla\cdot \bm u_{\rm f} = 0\,,
 \end{equation}
 together with non-slip boundary conditions at the surface of the
 particle and at the walls:
 \begin{eqnarray} \label{Eq2bc}
   && \bm u_{\rm f}(x,0,z) = \{V_{-},0,0\}\,,\quad \bm u_{\rm f}(x,H,z) = \{V_{+},0,0\}\,,\nonumber \\
   && \bm u_{\rm f}(x,y,z) = \bm u_{\rm p}(x,y,z)\ \ \mbox{at the
     particle's surface}.\nonumber
\end{eqnarray}
Here $\rho_{\rm f}$ is the carrier fluid density and $p$ is the
pressure. Clearly, on nano-scales the temperature difference (across a
particle) is sufficiently small to allow us to neglect the dependence of the
fluid's and particle's material parameters on the temperature, i.e. we take
$\nu$, $\chi_{\rm f}$ and $\chi_{\rm p}$ as constants.

The dynamics of a neutrally buoyant particle is governed by the
equations of the solid body rotation \cite{Allen}:
\begin{eqnarray}
  \frac{d \, {\Omega}_x^{(\mathrm{b})}  }{dt}& = &
  \frac{T_x^{(\mathrm{b})}}{I_x}
  +
  \frac{I_y-I_z}{I_x}\Omega_y^{(\mathrm{b})}\Omega_z^{(\mathrm{b})}\,, \nonumber\\
  \frac{d \,  {\Omega}_y^{(\mathrm{b})} }{dt} & = &
  \frac{T_y^{(\mathrm{b})}}{I_y}
  +
  \frac{I_z-I_x}{I_y}\Omega_z^{(\mathrm{b})}\Omega_x^{(\mathrm{b})}\,,  \label{Eq3}\\
  \frac{d\,  {\Omega}_z^{(\mathrm{b})}  }{dt} & = &
  \frac{T_z^{(\mathrm{b})}}{I_z}
  +
  \frac{I_x-I_y}{I_z}\Omega_x^{(\mathrm{b})}\Omega_y^{(\mathrm{b})}\,,\nonumber
\end{eqnarray}
\end{subequations}
where $\bm\Omega^{(\mathrm{b})}$ is the angular velocity in the
body-fixed frame, $\bm T^{(\mathrm{b})}$ is the torque in the
body-fixed frame and $I$ is the moment of inertia tensor.

\subsection{\label{ss:num} Numerical approach and its validation}
\begin{figure}[!t]
  \begin{center}
    \includegraphics[width=8.5cm]{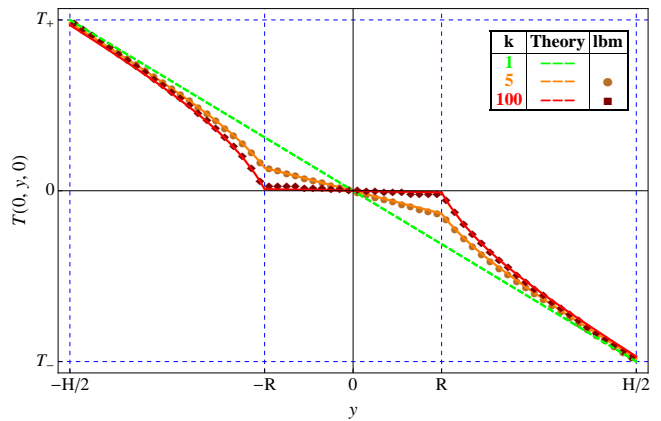}
  \end{center}
  \caption{\label{f:2} Temperature profile along a line passing
    through the center of a spherical particle (of radius $R$) in a
    quiescent fluid. Dashed (green) line -- conductive temperature
    profile in the pure fluid, solid (red) curve -- Eq.~(\ref{eq:LnL})
    \cite{LnL} for an infinite domain with a temperature gradient,
    dots -- numerical results (at lattice points) with $k = 5$ (full
    circles) and $k=100$ (squares). The present test was performed
    with Pr~$=1$, $H=64$ and $R=10$ lattice units.}
\end{figure}

The numerical simulation of the conjugated heat transfer problem
given by Eqs~\eqref{basicEqs} is performed by means of two coupled
D3Q19 Lattice Boltzmann (LB) equations under the so-called BGK
approximation \cite{Succi2001}. Details of the simulations are
presented in Appendix~\ref{ss:Numerics} . Besides purely numerical
means, the control of the simulations was done by its comparison with
known analytical solutions. As an example, Fig~\ref{f:2} shows the
comparison between the analytical temperature profile~(\ref{eq:LnL})
(solid red line) and numerical profile (black dots) for a moderately
big spherical particle [$H/(2R) = 3.2$].
The second test was the comparison in Fig.~\ref{Jeffery_Test} of the
particle's angular velocity theoretically obtained by Jeffery
\cite{Jef} and simulated by our LB-realization. For more details,
again see Appendix~\ref{ss:Numerics}.

\subsection{\label{volume}From single particle modeling to finite
  volume fraction}
Here we discuss how to relate the contribution of a single particle to
the heat flux with the total contribution of many {weakly-}interacting
particles randomly distributed in the flow occupying a finite volume
fraction $\varphi$.

{The first step is to consider \emph{fully dilute limit},
  $\varphi\to 0$, in which we can exploit the fact that} the particle
contribution is additive and proportional to $\varphi$.  {In
  this way} we can introduce the Nusselt number as the ratio between
the total heat flux (inside the computational cell) and the conductive
heat flux in the basic cell:
\begin{equation}
  \label{eq:NusseltDef}
  \mbox{Nu}= \frac{\langle q\rangle}{\langle q_{\rm cond}\rangle} = \frac{-\chi_{\rm eff}\nabla_{y} T}{-\chi_{f}\nabla_{y} T} = \frac{\chi_{\rm eff}}{\chi_{f}}\,,
\end{equation}
where $q$ is the total heat flux, $q_{\rm{cond}} = -\chi_{f}\nabla_{y}
T$ is the conductive part and $\chi_{\rm eff}$ is the effective heat
diffusivity of the composite fluid. Also, $\nabla_{y}T \equiv
{\partial T(x,y,z)}/{\partial y}$.  Without particles $\langle
q\rangle = \langle q_{\rm{cond}}\rangle$ and Nu~$=1$. For \emph{dilute
  suspensions}, in which $\varphi \ll 1$,  Nu can be expanded in
powers of $\varphi$~\cite{Jeff73}:
\begin{subequations}
  \label{53}
  \begin{eqnarray}\label{53A}
    \mbox{Nu} &=& 1+ \varphi\,K(\varphi)\,, \\ \label{53B}
    K(\varphi) &=& K_1   + K_2\, \varphi  +\dots\,,
  \end{eqnarray} 
\end{subequations}
where $K_i$ are dimensionless constants dependent on the Peclet
number, on the particle's aspect ratio, $r$, on the relative heat
diffusivity, $k$, and maybe other parameters like Reynolds or Prandtl
numbers [see Eqs.~(\ref{DimNumDefs})], i.e. $K_j = K_j(\mbox{Pe}, r,
k, \dots)$.

As expected, in our simulations the quantity $($Nu$-1)$ is indeed
proportional to $\varphi$ for $\varphi\lesssim 0.05$.  This allows us
to find $K_1$ as a function of other parameters~\eqref{DimNumDefs} of
the problem.

The next order term in the expansion~\eqref{53B}, i.e. $K_2\,\phi$, contains
very important information about how Nu depends on $\varphi$ for small
but finite values of $\varphi$. Generally speaking, to get this
information from numerics one needs to solve Eqs.~\eqref{basicEqs} for
many particles, {randomly} distributed in space.  We demonstrate
that this problem can be \emph{tremendously simplified} by reducing it
to a \emph{one-particle} case, in which this particle of volume
$V{_{\rm {p}}}=\varphi V{_{\rm {cell}}}$ is put in the center of
computational box of volume $V_{\rm {cell}}$ with \emph{periodic
  boundary conditions} in the horizontal (orthogonal to the
temperature gradient) $x$- and $z$-directions.  In this way the total
system can be considered as constructed from a periodic repetition of
the basic elementary cell in the $x$- and $z$-directions. Comparing in
Subsec.~\ref{sss:sphere-rest} our numerical results with analytical
findings obtained under the assumption of {random} particle distributions,
we see that the precise particle distribution is not essential -- what
really matters is the {actual} volume fraction.

The reason for that is quite simple:  as
one sees in Fig.~\ref{f:2}, the deviation from the linear temperature
profile becomes important at distances$\,$\footnote{It can be shown
  from Eq.~(\ref{eq:LnL}) that the distance should be smaller than $R
  \big| \delta^{-1} (k-1)/(k+2) \big|^{1/3}$, where $\delta$ is the
  deviation criterion, i.e. $\left| T_{\mathrm{f}}/T_{\mathrm{lin}} -
    1 \right| < \delta$ and $T_{\mathrm{lin}} = G y$.}  smaller than
the particle diameter $2R$ for both $k=5$ and $k=100$. Thus any nonlinear
dependence of Nu on $\varphi$ can appear only if the particles are sufficiently
close to overlap the deviation from linear temperature profile.  We see from our
numerics that the this interaction
causes 10\% deviation from the straight line in Nu vs. $\varphi$
dependence for volume fractions of about 0.1 and more for $k \leq
100$. For $\varphi>0.1$ there is some nonlinear dependence of Nu on
$\varphi$ but it practically coincides for the random and the periodic
distributions of particles. 

\section{\label{s:3Body}Towards an analytical model of the heat flux}
In this Section we begin with collecting the required information
about the dependence of $K_j$ on the various
parameters~(\ref{DimNumDefs}). To do this we consider simple limiting
cases, for which some of these parameters are put to zero. We start
with the case of spherical particles.

\subsection{Spherical nano-particles}
\begin{figure}
  \includegraphics[width=8.7cm]{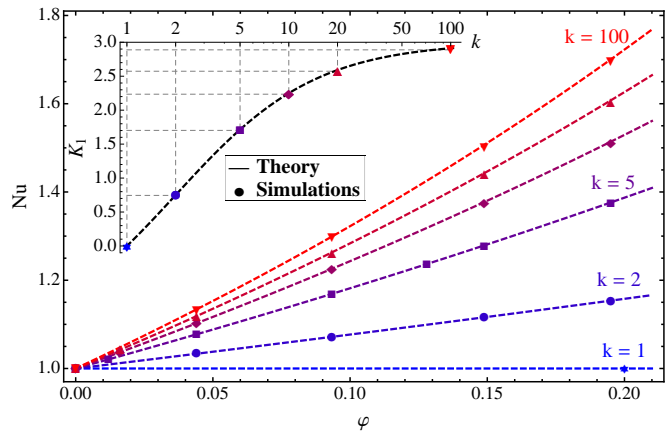}
  \caption{\label{fig:Nu64} (Color online). Quiescent fluid, spherical
    particle in the center.  Nu$(\varphi)$. Dashed curves -- theory,
    Eq.~(\ref{A-nlf}) and dots -- simulations for different $k$
    (color-coded from ``cold''-blue, $k=1$, to ``hot''-red, $k=100$).
    Insert: $K_1(k)$. Dashed curve -- theory, Eq.~(\ref{A-nlf}) and
    dots -- simulations for different $k$ (same color-code). {Notice,
      that for $128^3$ runs the accuracy is better then $0.3$\%, while
      for $64^3$ runs it is about $0.7$\%, and only for really small
      $32^3$ runs it reaches the worst value of $1.5$\%. These means
      that for our purposes, usually the runs with $64^3$ box is fully
      satisfactory.}}
\end{figure}
\begin{figure*}
  \begin{center}
    \includegraphics[width=8.4cm]{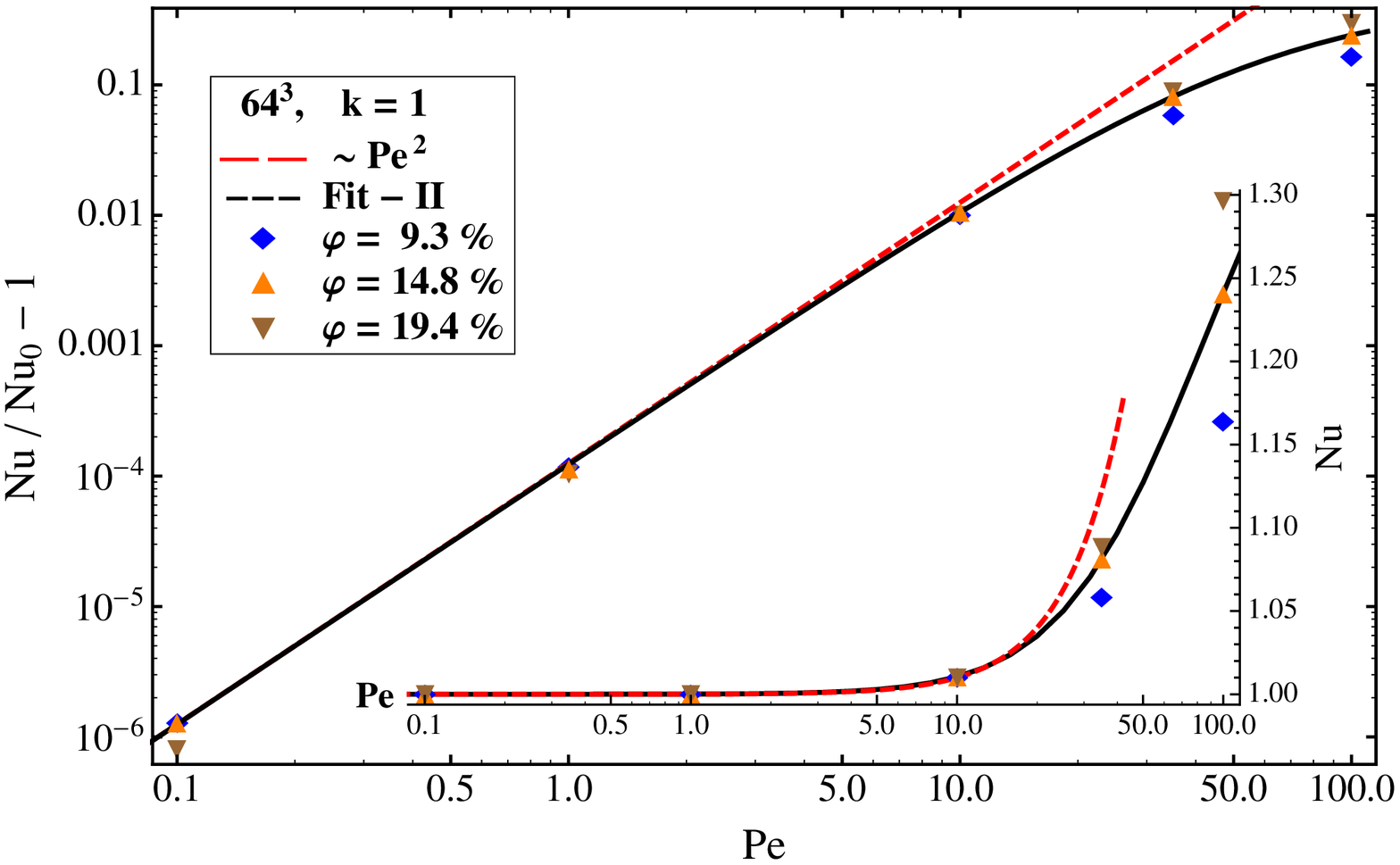}
    \includegraphics[width=8.2cm]{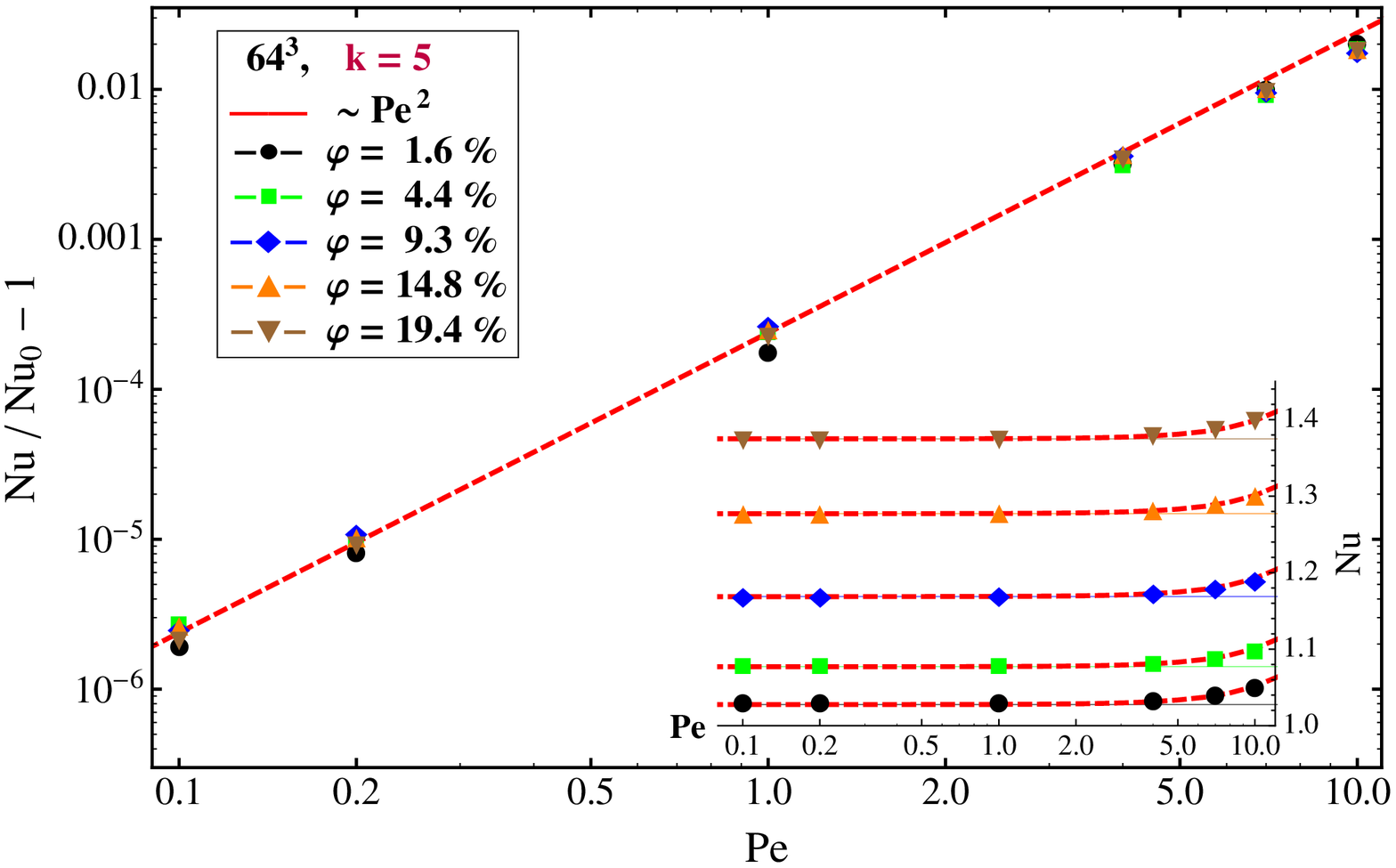}\\
    \includegraphics[width=8.2cm]{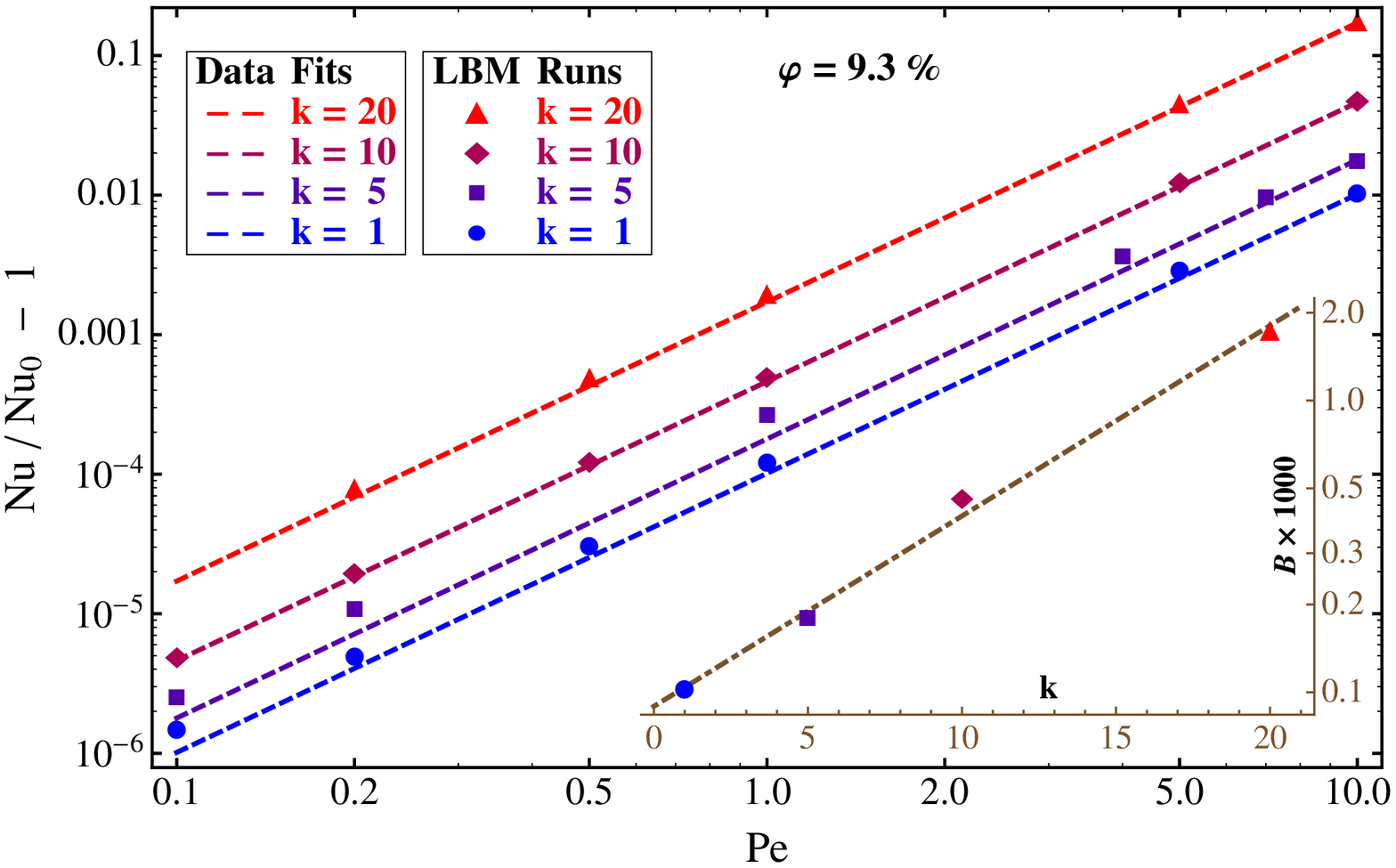}
    \includegraphics[width=7.9cm]{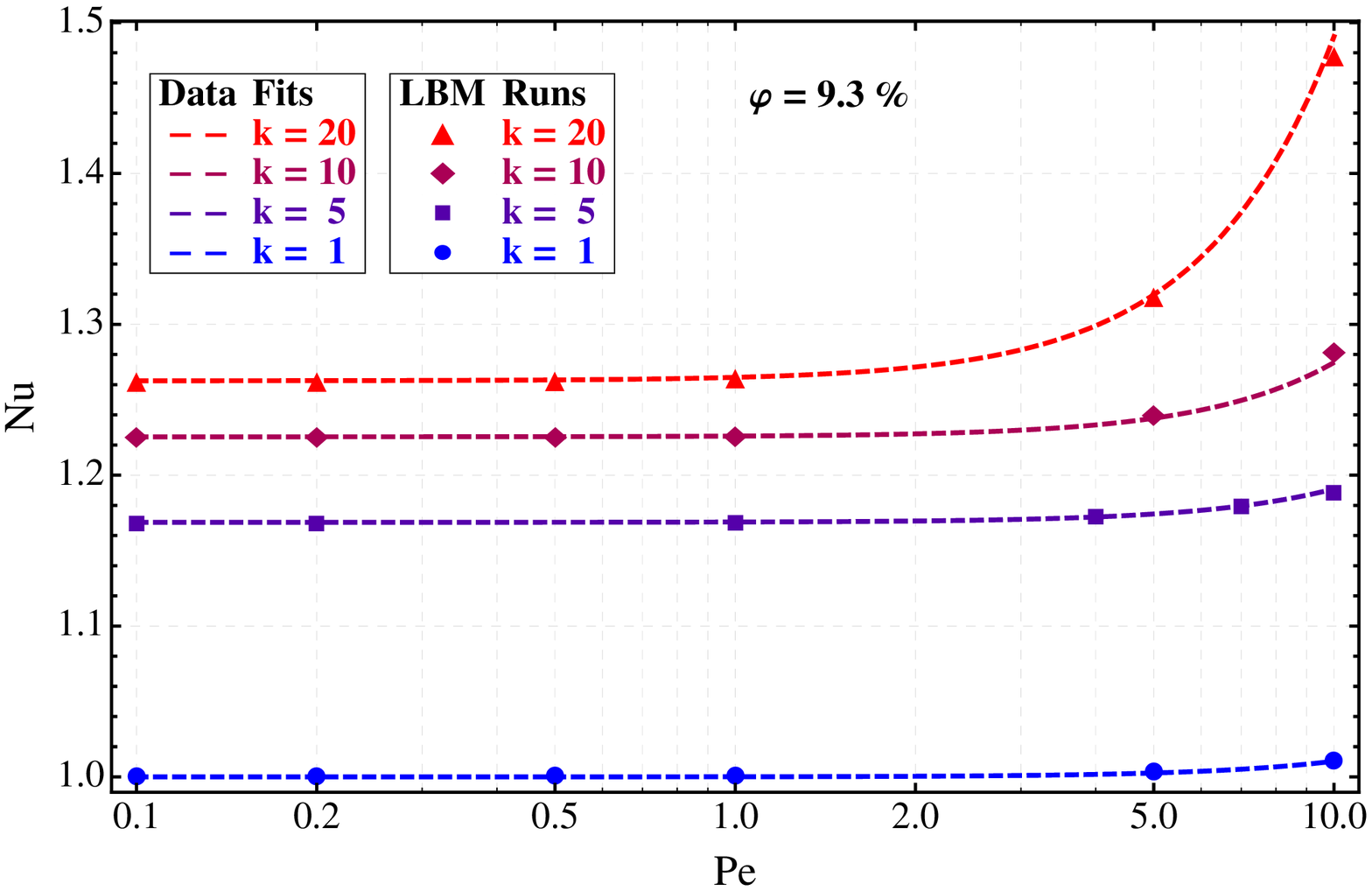}
  \end{center}
  \caption{\label{f:4} (Color online).  A spherical particle in a
    shear (Couette) flow.  \textbf{Upper left panel:} $k = 1$,
    $\varphi=9.3\%, 14.8\%$ and 19.4\%.  \textbf{Upper right panel:}
    $k = 5$ for wider region of $\varphi$, but Pe is limited by 10.
    Both upper panels show Nu(Pe)/Nu$_0-1$ \textit{vs}. Pe, and both
    insets show Nu(Pe) \textit{vs}. Pe. Here Nu$_0
    \equiv\,$Nu$(0)$. Dashed (red) line is a fit $\propto$ Pe$^2$,
    cf. Eq.~(\ref{Nu-Pe}); (color) symbols are simulations with
    different volume fractions: $\varphi=$ 1.6\%, 4.4\%, 9.3\%,
    14.8\%, 19.4\%. Solid (black) curve on upper left panel represents
    the fit by Eq.~(\ref{largePe}).  \textbf{Lower panels:}
    $\varphi=9.3\%$ and different $k$ (from ``cold''-blue, $k=1$, to
    ``hot''-red, $k=20$).  Dashed (''temperature''-colored) lines are
    fits to simulated data (symbols, same colors) by
    Eq.~(\ref{Nu-Pe}), left panel, and by combined
    Eqs.~(\ref{eq:BvsK}) and (\ref{A-nlf}), right panel.
    \textbf{Inset on lower left panel}: $B(k,\varphi)$ \textit{vs} $k$
    at $\varphi=9.3\%$. Dot-dashed (brown) line is the fit to
    Eq.~(\ref{eqsForB}).  Parameters: $64^3$, Pr~$= 1$, thus,
    $\mathrm{Re} = \mathrm{Pe}\,k$.  }
\end{figure*}

\subsubsection{\label{sss:sphere-rest} Test case: s spherical
  {nano}-particle in a fluid at rest}

{Here we consider the simplest case of} a spherical particle ($r=1$)
in a fluid at rest (Re = Pe = 0).  Chiew and Glandt~\cite{chiew}
suggested the following formula for this case: {\begin{subequations}
    \label{A-nlf}
    \begin{eqnarray}
      K(k,\varphi) &\simeq& \frac{3\, K_1(k)}{3-K_1(k)\varphi}\,, \quad K_1(k) = 3\,\frac{k-1}{k+2}\,,~~~~ \\
      \mbox{Nu}&=&  1 +  \frac{3\, K_1(k)\, \varphi}{3-K_1(k)\, \varphi}\ .
    \end{eqnarray}
\end{subequations}
We notice that the expression for $K_1(k)$ which controls the very
dilute limit ($\varphi\to 0$) is the same as in the Maxwell model
\cite{Max}. Obviously, for $k=1$ one has zero effect,
i.e. Nu~$=1$. For $k\to \infty$, $K_1=3$ and one has maximal possible
enhancement (at fixed $\varphi \ll 1$).  For $k=2$ one has 25\% of
this value, $k=4$ gives already 50\% of the effect, $k=10 \Rightarrow
75\%$, $k=20 \Rightarrow 86\%$ and $k=100\Rightarrow 97\%$. Clearly,
the larger $k$ the better, but increasing $k$ above 100 is not
effective.

Actually, Eq.~(\ref{A-nlf}) includes also information about the
nonlinear dependence $K(\varphi)$. However this dependence is very
weak in the relevant range of parameters: e.g. for $k=10$ and
$\varphi=0.2$ the difference between $K(\varphi)$ and its linear
approximation is only 15\%. For smaller $k$ this difference is even
smaller. The physical reason for that was explained at the end of previous
Section.

We tested the prediction of Eq.~(\ref{A-nlf}) by simulating the heat
flux in a periodic box as explained above in Sec.~\ref{s:model} and
in Appendix~\ref{ss:Numerics}. For the fluid at rest the (volume
averaged) Nusselt number (as a function of $\varphi$ at different
values of $k$) was measured and is shown in Fig.~\ref{fig:Nu64}, where
the inset shows $K_1(k)$.  The overall conclusion is that

\vskip 0.1 cm
\noindent \emph{Eq.~(\ref{A-nlf}) agrees extremely well with all our
  simulations and thus can be used in modeling the effect of spherical
  particles in a fluid at rest.}

\subsubsection{\label{SphereShearFlow}Spherical nano-particle in a shear   flow}}
The next important question is how the particle rotation affects the
heat flux.  As already mentioned, this rotation induces fluid motions
around the particles thus causing convective contribution to the heat
flux. This changes the temperature profile around the particles and,
in turn, affects the heat flux inside the particles. To study these
issues analytically is extremely difficult and thus numerical
simulations can play a crucial role. Due to the fact that we have to
deal with a number of parameters, we will carefully examine them
starting with the simplest case of spherical particles. Results for
spheroidal particles are discussed in Sec.~\ref{sss:el-rot}.

We begin by considering the case $k=1$; such a particle at rest does
not affect the heat flux, thus Nu~$=1$. In Fig.~\ref{f:4}, upper left
panel, we present Nu(Pe)/Nu$(0)-1$ as a function of Pe for $k=1$ at
various volume fractions (from {9.3\%} to 19.4\%), while the inset is
showing Nu \textit{vs}. Pe.  Observing the data, {we suggest for
  Pe$\lesssim 10 $ the following ansatz}
\begin{subequations} \label{Pe^2} \begin{equation}
    \label{Nu-Pe}
    \mbox{Nu(Pe)}\simeq \mbox{Nu}(0)[1+B(k,\varphi)\mbox{Pe}^2] \,,
\end{equation}
{shown in this panel as a straight dashed (red) line. One sees
  that Eq.~(\ref{Nu-Pe}) is approximately valid up to Pe~$\simeq 10$,
  where the deviation is about 1\%.  We note also that for $k=1$ the
  value of $B$ is very small, $B(1,\varphi)\simeq 10^{-4}$.}  A
similar analysis for $k=5$, (cf.  Fig.~\ref{f:4}, upper right panel),
shows that Eq.~(\ref{Nu-Pe}) fits the data even better with a similar
value $B(5,\varphi)\simeq 2.5\cdot 10^{-4}$.

We see that $B$ depends weakly on $\varphi$ but more strongly on
$k$. Thus, to determine the leading $k$-dependence we put a sphere of
radius 18 in a $64^3$ computational volume (all in lattice units);
this is equivalent to $\varphi \simeq 9.3 \%$, see Fig. \ref{f:4},
lower panels. The $k$-dependence of $B(k,\varphi)$ in
Eq.~{(\ref{Nu-Pe})} in the range of $\varphi\simeq 10\%$ can be fitted
by:

\label{eq:BvsK}
\begin{eqnarray}
	\label{eqsForB}
  B(k,\varphi) &\simeq& B_1(\varphi)\, \exp\left[B_2(\varphi)\, k \right] , \\
  B_1 &\simeq& 9 \!\times\! 10^{-5}, \ B_2 \simeq 0.15\ .
	\end{eqnarray}
\end{subequations}

For larger Pe$\gtrsim 10 $ , the Nu~\emph{vs.}~Pe dependence deviates
down, as expected, because at Pe~$\to \infty$ this dependence should
saturate. Our analysis shows that this dependence can be fitted
  with a good accuracy by the formula that generalizes
  Eq.~\eqref{Nu-Pe}:
  \begin{equation}
  \label{largePe}
  \frac{\mbox{Nu(Pe)}}{\mbox{Nu}(0)} - 1 \simeq
 \frac{B(k,\varphi)\ \mbox{Pe}^2}{1+  \mbox{Pe}/\mbox{Pe}_1(k)+ [\mbox{Pe}/\mbox{Pe}_2(k)]^2}\ .
\end{equation}
The upper left panel in Fig.~\ref{f:4} shows by a solid (black) curve how this model
works for $k=1$ (with Pe$_1(1)\approx 61$ and Pe$_2(1)\approx 63$).

Having in mind that in many applications {related to nanofluids} the value of Pe is smaller
than 0.01 and in any case rarely exceeds unity, we reach the
conclusion that the convective heat flux around spherical particles
and variations of the heat flux inside spherical particles due to
their rotation can be neglected. Equation~(\ref{A-nlf}) can be used to
model the heat flux enhanced by spherical nano-particles with finite
volume fraction up to 25\% and any actual value of $k$ in fluids at
rest and in shear flows.

The next question to consider is the effect of the particle shape. We
will begin with the case of spheroidal particles in fluids at rest.

\begin{figure}
  \begin{center}
    \includegraphics[width=8.4 cm]{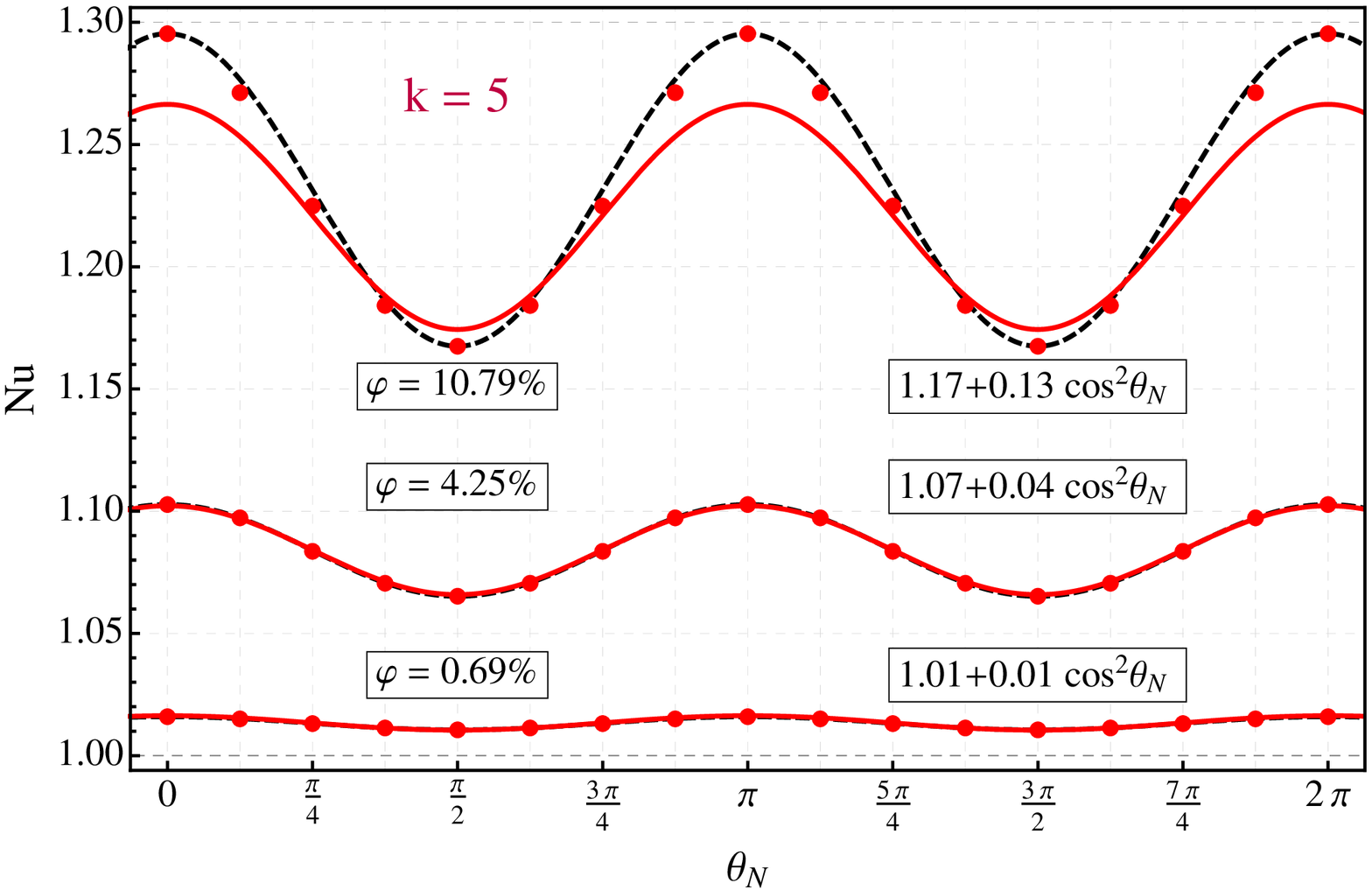}
     \includegraphics[width=8.4 cm]{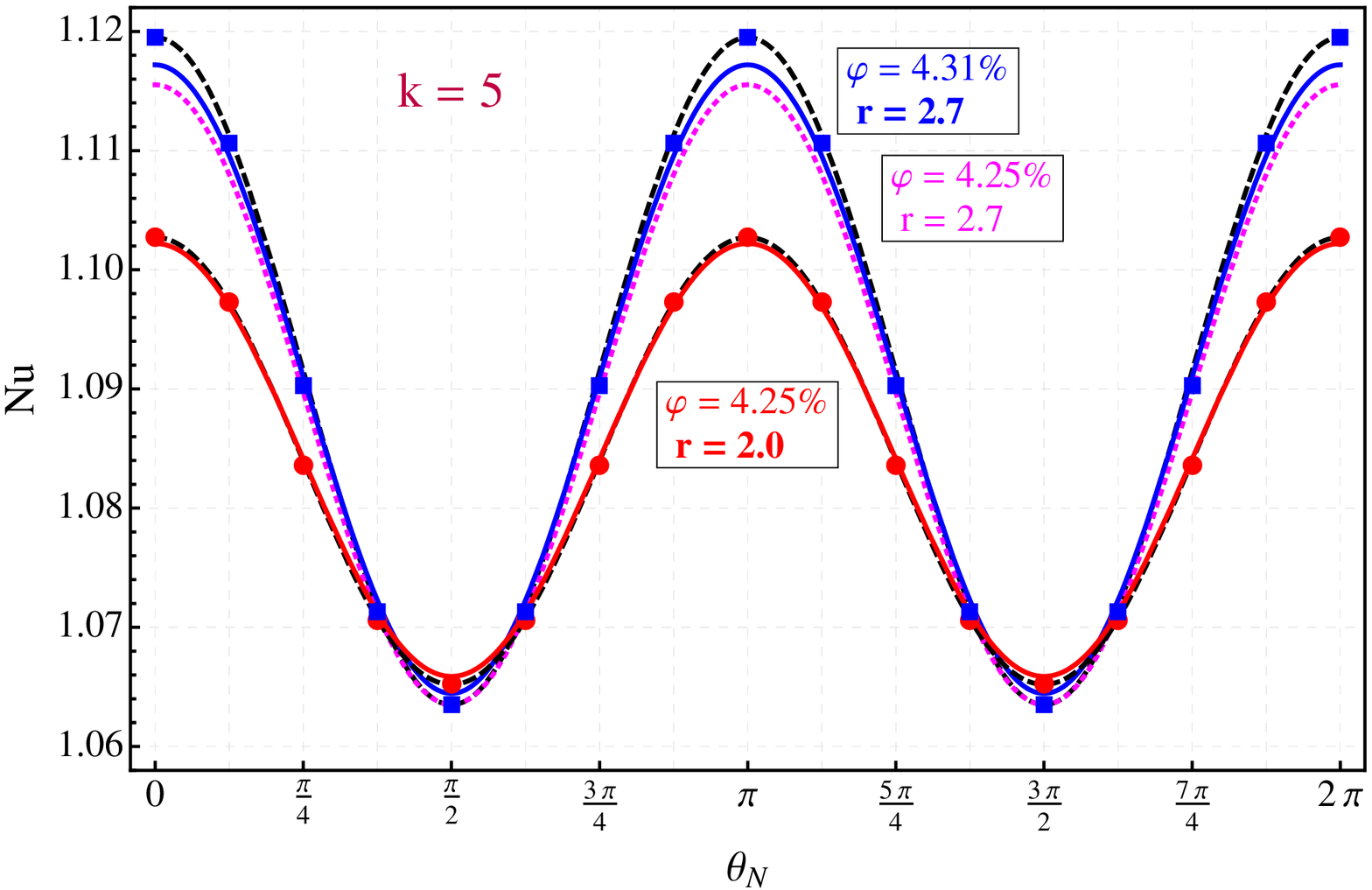}
  \end{center}
  \caption{	\label{NuVsTheta} (Color online)
	Spheroid in a quiescent fluid at different angles $\theta_{\rm N}$. Both panels: $k=5$, Pr = 1, $64^3$. Symbols -- simulations, solid (red or blue) curves -- Eq.~(\ref{spheroid}) with appropriate $r$ and $\varphi$, dashed (black) curves -- fits by Eq.~(\ref{NuSpheroisBoth}) to the simulations.
	\textbf{Upper panel:} $r=2$. Data: $\varphi \simeq 10.79\%$ (upper), $\varphi \simeq 4.25\%$ (middle) and $\varphi \simeq 0.69\%$ (lower). The largest mismatch between the simulations and Eq.~(\ref{spheroid}) is 2.8\% at $\varphi = 10.79\%$.
	\textbf{Lower panel:} lower (red) dots: $\varphi = 4.25\%$, $r=2.0$,  and upper (blue) squares: $\varphi = 4.31\%$, $r=2.7$. The dotted (magenta) curve is Eq.~(\ref{spheroid}) with $\varphi = 4.25\%$ and $r=2.7$. The effect of changing $r$ is larger than that of $\varphi$.
  }
\end{figure}

\subsection{\label{ss:spheroids-F} Spheroidal nano-particles}

\subsubsection{\label{sss:el-fix}Spheroidal nano-particle in a fluid at rest}

The general expectation is that spheroidal nano-particles should be able to enhance the heat flux much
more effectively then spherical particles. To achieve this they have to
be oriented in the ``right'' direction, i.e. along with the temperature
gradient. Therefore, logically, the study of the heat flux in
nanofluids laden with spheroids should begin with the clarification of the
effect of the spheroid orientation on the heat flux. In this Section
we consider this effect for a given and fixed orientation of the
spheroids. For this purpose we perform numerical simulations of the
heat flux with spheroidal nano-particles in a quiescent fluid (Pe$ =
0$) with a temperature gradient, see Fig.~\ref{geometry}, taking for
concreteness $k=5$.  

For different orientations of the spheroid
(i.e. different $\theta_{\rm N}$ -- the angle between the temperature
gradient vector and the largest spheroid's axis), we measure different
Nusselt numbers: when $\theta_{\rm N} = 0$ spheroids with higher
conductivities tend to forms a thermal shortcut, thus Nu is larger
than for $\theta_{\rm N} = \pi/2$, where this shortcut effect is
reduced (Fig.~\ref{NuVsTheta}). Our numerics shows that the dependence
of the Nusselt number on the spheroid orientation,
i.e. Nu$(\theta_{\rm N})$, for small $\varphi$ is well described by a
simple formula
\begin{subequations}\label{K}
\label{NuSpheroisBoth}
  \begin{eqnarray}
    \label{NuSpherois}
    \mbox{Nu} &=& 1+\varphi\, \mathcal  K(k,r,\varphi,\theta_{\rm N})\,,   \\
    \label{cos2}
    \mathcal  K(k,r,\varphi,\theta_{\rm N}) &=& \mathcal  K_{\rm min} +  \mathcal  K_{\rm amp}\,  \cos^2(\theta_{\rm N})\,,
\end{eqnarray}
\end{subequations}
and both $\mathcal  K_{\rm min}$ and $\mathcal  K_{\rm amp}$ are functions of $k, r$
and $\varphi$. This equation is a generalization of Eq.~(\ref{53}) for
the case of spheroids, and it was suggested in~\cite{CosSquare} for
the limiting case of $r \to \infty$ ({nanotubes}).  For large $\varphi> 5\%$ deviations become visible. For
finite $\varphi$ there have been attempts to study the role of the
particles shape and form-factors, e.g. by Nan et al. \cite{97Nan}:
\begin{eqnarray}
  \label{spheroid}
  \label{Nan-form-factor}
  \mathcal  K &=&  \frac{\beta_1 +(\beta_2 -\beta_1)\langle \cos^2 \! \theta_{\rm N} \rangle}{1 -\varphi \left[L_1 \beta_1 +(L_2 \beta_2 -L_1 \beta_1)\langle \cos^2 \! \theta_{\rm N} \rangle \right]}\,,~~~~ \\
  \label{Nan-form-factor-b}
  \mbox{where:} \!\! && \!\! \beta_j = \left[{L_j + 1/(k-1)}\right]^{-1}, \quad j = 1,2\,, \nonumber\\
  && L_2 = 1 -2 L_1\,, \nonumber \\
  \label{Nan-form-factor-c}
  \mbox{$r > 1$:} \!\! && \!\! L_1 = \frac{r}{2(r^2-1)}\left[ r -\frac{\textrm{arccosh}(r)}{\sqrt{r^2-1}} \right]\ .\nonumber
\end{eqnarray}
Here $\langle\dots\rangle$ stands for an ensemble
average. Clearly, in the case of a single particle as in our
  simulations, or a periodic array of the particles, the average
  coincides with the single value.  Notice, {also} that for
spheres, i.e. when $r\rightarrow 1$, $L_1 = L_2 = 1/3$, $\beta_1 =
\beta_2 = K_1$ from Eq.~(\ref{A-nlf}), and Eq.~(\ref{Nan-form-factor})
simplifies to Eq.~(\ref{A-nlf}). For small $\varphi$
Eq.~(\ref{Nan-form-factor}) coincides with Eq.~(\ref{cos2}), giving
analytical expression for $\mathcal  K_{\rm min}$ and $\mathcal  K_{\rm amp}$.

The range of validity of equation (\ref{Nan-form-factor}) was tested
by means of a series of simulations at varying angles and volume
fractions. The results are reported in Figure \ref{NuVsTheta}. We can
conclude that the model in Eqs.~(\ref{spheroid}) well represents the
heat flux in the presence of resting spheroidal particles in the
relevant range of the parameters $k$, $r$ and $\varphi$.  To apply
Eqs.~(\ref{spheroid}) for spheroids in a simple shear flow we have to
find how the ensemble average $\langle \cos^2 \theta_{\rm N} \rangle$
depends on $r$. For this purpose we first need to know the
orientational distribution function of spheroids in the shear
flow. This is a subject of the following Subsection.

\subsubsection{\label{sss:el-rot}{Spheroidal nano-particle in a shear flow}}

\begin{figure}[h] \vskip -0.24cm
  \includegraphics[width=8.8 cm]{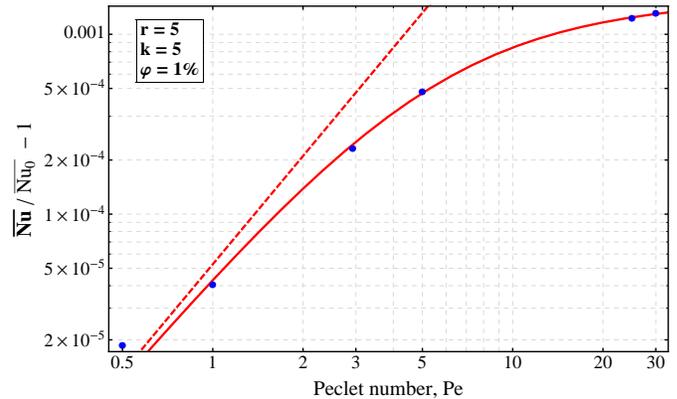}
  \caption{\label{f:9} (Color online). Example of advanced
    fit~\eqref{largePe} for a spheroid $r=5$, $k=5$. (Blue) dots --
    numerical data, solid (red) curve -- fit by
    Eq.~fit~\eqref{largePe} with $B\approx 5.3 \times 10^{-5}$,
    Pe$_1\approx 5.2$ and Pe$_2\approx 30.3$. Dashed (red) line is
    $B\,\mathrm{Pe}^{2}$.  Here $\overline{\mathrm{Nu}}$ is the
      time-averaged Nu$(t)$ over the period of rotation (determined in
      the simulations), and $\overline{\mathrm{Nu}_0}$ is the
      time-averaged Nu$(t)$ for smallest available $\mathrm{Pe}$,
      e.g. $\mathrm{Pe} = 0.01$.  }
\end{figure}

The effect of rotation of elongated spheroidal particles ($r>5$) is very
similar to that of spherical ones ($r=1$), just the parameters $B$,
Pe$_1$ and Pe$_2$ of the advanced fit~\eqref{largePe} depend on the aspect
ratio $r$. As an example, we presented in Fig.~\ref{f:9} a
  preliminary result of the computed and fitted Nu(Pe) dependence for
$r=5$ and $k=5$. In this case $B\approx 5.3 \times 10^{-5}$,
Pe$_1\approx 5.2$, Pe$_2\approx 30.3$. This is interesting to compare
with the respective parameters for $r=1$ and $k=5$: $B\approx 2.5 \times
10^{-4}$, Pe$_1\approx 61$, Pe$_1\approx 63$.  One sees that the
Pe-enhancement of the heat flux with elongated particles is smaller
then for spherical ones [$B (r>1)< B(r=1)$] and it saturates at smaller
Pe: Pe$_1(r>1)<$Pe$_1(r=1)$ and Pe$_2(r>1)<\mathrm{Pe}_2(r=1)$.
Moreover, the overall conclusion that Nu(Pe) is almost
  Pe-independent for Pe$\lesssim1$ remains valid, thus, the model
  developed above in Eq.~(\ref{spheroid}) with Eq.~(\ref{cos-2}) may
  be safely used for nano-particles laden flows at Pe$\lesssim1$.

\section{\label{s:4Model}Analytical model of heat transfer in a
  laminar shear flow}
In this section we develop an analytical model for $\mathrm{Nu}(k,r,\varphi)$.

\subsection{\label{s:stat}{ Orientational statistics of elongated
    nano-particles}}
\subsubsection{\label{sss:FPE} { Fokker-Plank equations for orientational PDF }}

In order to study the orientational statistics of elongated
nano-particles we consider the Probability Distribution Function
(PDF), $\mathcal P (\theta,\phi,t )$, which is the probability $P
(\theta,\phi,t )$ of finding any particular spheroid with its axis of
revolution in the interval $[\theta,\theta+d\theta]\times
[\phi,\phi+d\phi]$ on the unit sphere. {The PDF is then defined by}
\begin{equation}\label{P}
  P (\theta,\phi,t ) = \mathcal  P (\theta,\phi,t ) \sin\theta d\phi\,  d\theta \ .
\end{equation}
It was shown by Burgers \cite{Burgers1938} that $\mathcal P
(\theta,\phi,t )$ satisfies a generalized Fokker-Planck equation in
the presence of a shear:
\begin{equation}\label{FP}
  \frac{\partial \cal P}{\partial t} = - \bm  \nabla \cdot (\bm  w \mathcal  P)+ D_r \nabla^2 \mathcal  P ,
\end{equation}
where $\bm w\equiv (0,\dot \theta, \dot\phi \sin\theta)$ is the
relative velocity on the unit sphere of the axis of revolution for a
particle with instantaneous orientation $(\theta,\phi)$ ignoring all
Brownian effects.  The explicit form of this equation in spherical
coordinates is:
 \begin{subequations}\label{PDFeq}
   {\begin{equation}\label{PDFeq-a} {\frac{ \partial \mathcal
           P(\theta,\phi)}{\partial t} + \frac1{\sin\theta}\left[
         \frac\partial {\partial\theta} F_\theta \sin\theta +
         \frac\partial{\partial\phi}F_\phi\right]=0 \ .}\end{equation}}%
  Here $F_\theta$ and $F_\phi$, the $\theta$ and $\phi$
   projections of the {probability density} fluxes that have two
   shear-induced contributions, one proportional to $w$ and the other
   to the rotational diffusion coefficient, $ D_{\rm {r}}$:
\begin{eqnarray}\label{PDFeq-b}
{{F_\theta= \Big [  \frac{\partial \theta}{\partial t}-D_{\rm {r}} \frac{\partial}{\partial \theta} \Big ]
\mathcal{P}(\theta,\phi)\,,}}~~~~~~~~~~~
\\
{{F_\phi= \big [\sin \theta\, \frac{\partial \phi}{\partial t}- \frac{D_{\rm {r}}}{\sin \theta} \, \frac{\partial }{\partial \phi}\big ]\mathcal{P}(\theta,\phi)\ .}}
\end{eqnarray}\end{subequations}
Burgers in Ref.~\onlinecite{Burgers1938} derived the rotational
diffusion coefficient, $D_{\rm {r}}$, of elongated rigid spheroids of
revolution:
\begin{subequations}\label{dc}
\begin{eqnarray}
	D_{\rm {r}} &=& \frac{k{_{_{\rm {B}}}} T}{ 4 \mu V } \Big [\frac {r^2+1}{r^2 K_3 + K_1} \Big ]^{-1} \ .
\end{eqnarray}
Here $\mu$ is the dynamical fluid viscosity, $k{_{_{\rm {B}}}}= 1.374 \times
10^{-16}$erg/deg K is the Boltsmann constant, particle volume $V=\frac
43 \pi a^2 b$, $2a=d$, $b = r a$, $r \geq 1$ and
\begin{eqnarray} 
    K_3&=& \int_0^\infty \frac{r dx} {(r^2+x)^{3/2}(1+x)} \\ \nonumber
     &=&\frac{2}{r^2-1}\left[ 1 -\frac{r\ \mathrm{arccosh}(r)}{\sqrt{r^2-1}} \right]   \rightarrow  2\frac{\ln (2\,r/e)}{r^2}  \,, \\
 K_1&=& \int_0^\infty \frac{r dx} {(r^2+x)^{1/2}(1+x)^2} ~~~~~~~~~~  \\ \nonumber 
 &=& \frac{r}{r^2-1}\left[ r -\frac{\mathrm{arccosh}(r)}{\sqrt{r^2-1}} \right]
 \rightarrow 1 \ .
\end{eqnarray}\end{subequations}
This asymptotics is formally valid for $r\gg 1$, but give better than
10\% accuracy already for $r > 4$, allowing us to suggest the
approximate ``practical'' formula
\begin{equation}
	\label{D-approx}
	D_{\rm {r}}\simeq \widetilde D_{\rm {r}} = \frac{k{_{_{\rm {B}}}} T}{ 4 \mu V }\frac{\ln (4 r^2/e)}{(r^2+2\, r/ 3 -1)}\,,
\end{equation}
that works with an accuracy of about $10\%$ for any $r\ge1$ and better than
with $5\%$-accuracy for $r>10$.

The complete analysis of Eqs.~(\ref{PDFeq}) is very involved, see
e.g. Refs.~\cite{Leal1971} and \cite{Hinch1972}. We consider only the
small-diffusion limit, see the next subsection.

\subsubsection{\label{ss:small} {Statistics of elongated     nano-particles in the small diffusion limit }}

Jeffery~\cite{Jef} has shown that if inertial and Brownian motion
affects are completely neglected, then the motion of the axis of
revolution of a spheroidal particle is described by
\begin{subequations}
  \label{JeffOrbitsEqs}
  \begin{eqnarray}
    \tan \phi &=& r \tan \tau\,, \\
    \label{JeffOrbitsEqsTheta}
    \tan{ \theta} &=& C \sqrt{\cos^2 \tau +r^2 \sin^2  \tau} \\
    \nonumber
    &=& C\,r  \left( r^2 \cos^2\phi + \sin^2\phi \right)^{-1/2}, 
  \end{eqnarray}
\end{subequations}
where $\tau = t/T_{\rm r}$ with $T_{\rm r} = (r +1/r) /S$, and the
constant of integration $C$ is called the (Jeffery) orbit constant.

To analyze the small-diffusion limit we introduce two time-scales. The
first one defines the periodic motion that a nano-particle with a
finite $r$ exhibits: $T_{\rm r} = T_{\rm J}/2\pi$, where $T_{\rm J} =
2\pi (r +1/r) /S$ is the Jeffery's period. The second time scale is
determined by the inverse shear, $T_S=S^{-1}$.  Clearly, for large $r$
this is a much shorter time scale, so for $r \gg 1$,
\begin{equation}
  \label{slow}
  T_r\approx r/S=r\, T_S \ .
\end{equation} 
It means that the particles spend most of their time near pre- and
post-aligned states. If the Brownian diffusion is small enough such
that $T_{\rm dif} = D_\mathrm{r}^{-1} \gg T_r$ one can neglect the
effect of the Brownian motion on the dynamic motions of particles
along Jeffery orbits \cite{Jef} even during their slow time
evolution. In this case the stationary Fokker-Plank Eq.~(\ref{FP})
takes the simple form
\begin{equation}\label{FP-1a}
  \bm  \nabla \cdot (\bm  w \mathcal  P) =0\ .
\end{equation}
This equation can be solved {\cite{Leal1971, Hinch1972}}, giving
\begin{subequations}\label{PDF1} {\begin{eqnarray}
      \mathcal  P(\theta,   \phi) &=&  {f(C,r)}\, \mathcal  P_C(  \theta,   \phi)\,, \\
 	\label{FP-2a}
	\mathcal  P_C(  \theta,   \phi) &=& \left( {\! r \, \sin   \theta \cos^2\! \theta \sqrt{\cos^2\!\phi + \sin ^2\!\phi /r^2\  }} \right)^{\!\!-1}\!\!\!,~~~~~~~
      \end{eqnarray}}\end{subequations}
  where $\mathcal  P_C(  \theta,   \phi)$ is the PDF along a particular Jeffery orbit  with a given integration constant $C$, and {${f(C,r)}$} is the probability to occupy this orbit. This function is normalized as follows:
$$ \int_0^\infty f(C)dC = \frac 1{4\pi}\ .
$$
{${f(C,r)}$} was found in {\cite{Leal1971}} for the limiting
cases:
\begin{subequations} \label{f}
  \begin{eqnarray}
    f(C,r  =1) &=& \frac{1}{4\pi}\frac{C}{(C^2+1)^{3/2}} \,, \label{r=1}   \\
    f(C,r \gg 1) &\simeq& \frac{1}{\pi}\frac{C}{(4C^2+1)^{3/2}} \label{r=100}\,, \\
    f(C,r \ll  1) &\simeq& \frac{1}{\pi}\frac{C r^2}{(4 r^2 C^2+1)^{3/2}} \label{r=0}\ .
  \end{eqnarray}  \end{subequations}
Note that Eq.~(\ref{r=1}) is exact, providing a consistency check of
the present approach by comparison with spherical particles. The
asymptotic, Eq. (\ref{r=100}), is very accurate for $r>20$, but already
for $r \gtrsim 4$ it provides reasonable accuracy (better then $\sim
10\%$).

\begin{figure}[t]
  \begin{center}
    \includegraphics[width=8.4cm]{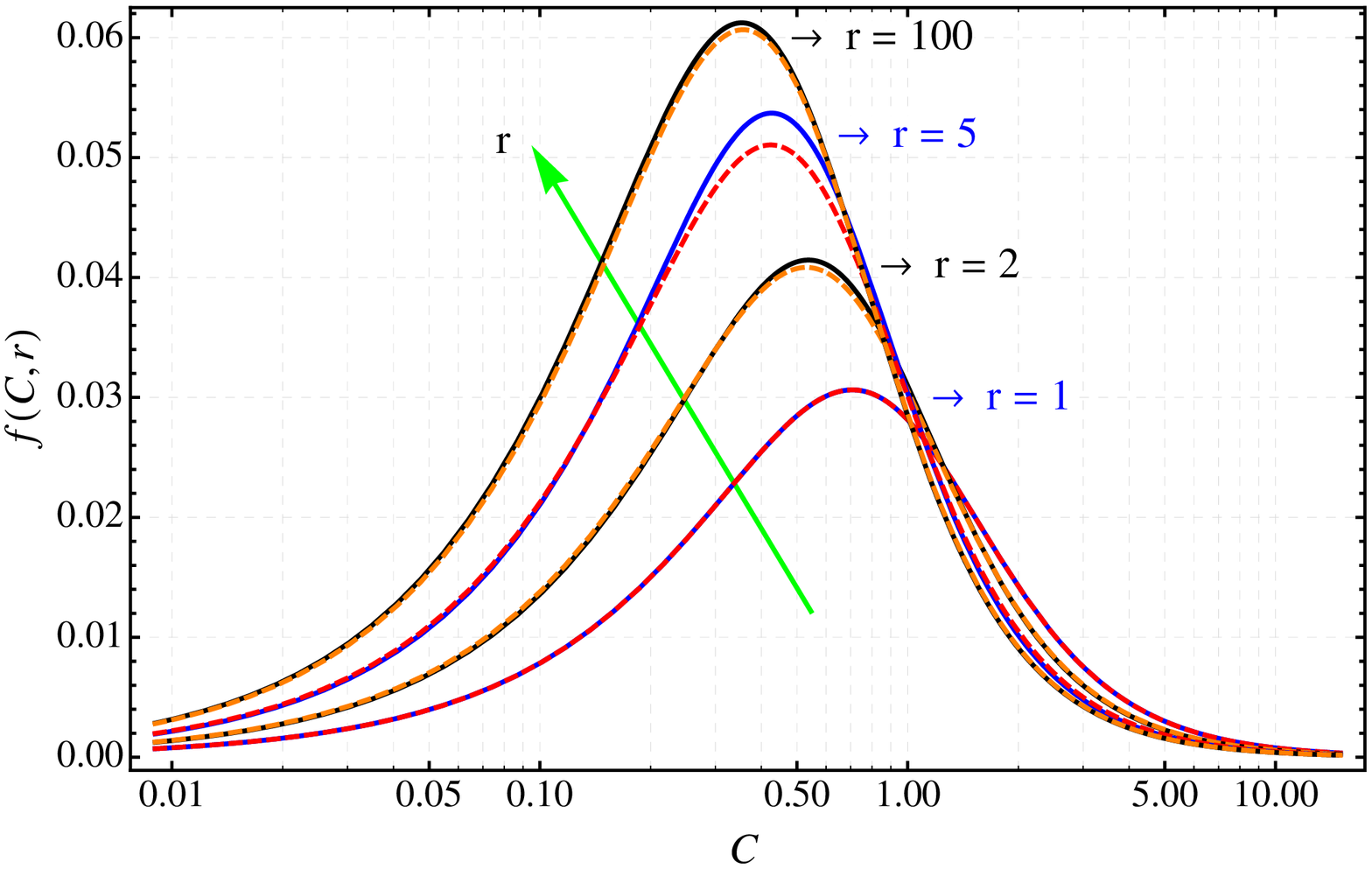}\\
    \includegraphics[width=8.4cm]{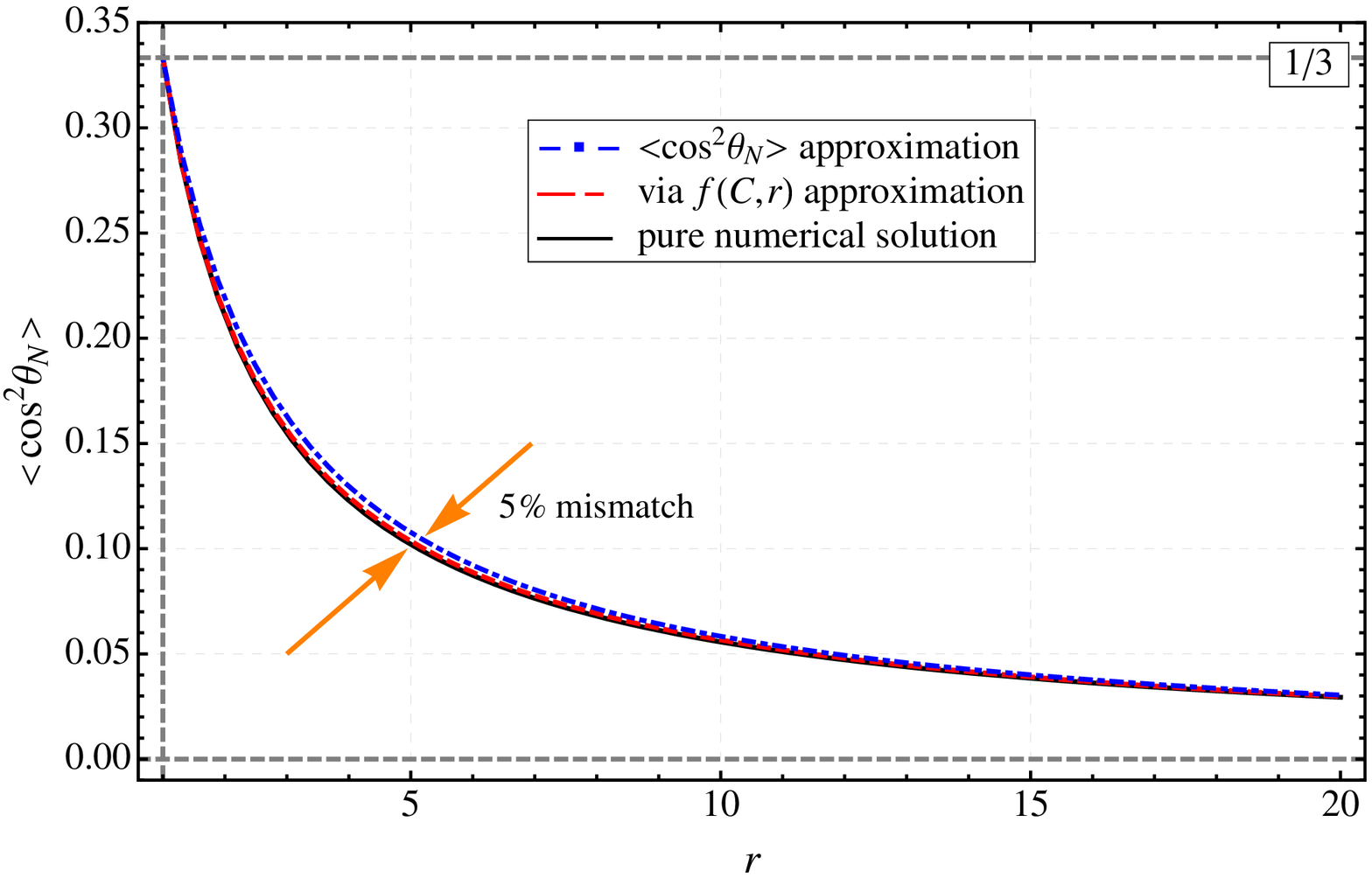}
  \end{center}
  \caption{\label{fig:PDFs} (Color online). \textbf{Upper panel:}
    Comparison of ``exact numerical", solutions, obtained from
    Eq.~(\ref{FP-1a}) (solid lines) and approximate solutions
    Eq.~(\ref{approx-f}) for different $r$ (dashed lines). There is
    visible difference (about 5\% in the value of maximum) only for
    $r=5$.  \textbf{Lower panel:} Comparison of dependence of $\langle
    \cos^2 \theta \rangle$ vs. $r$ obtained by a) applying ``exact
    numerical" PDF (black solid line), b) approximate analytical PDF,
    Eq.~(\ref{approx-f}), (red dashed line) and c) analytical
    approximation for this dependence Eq.~(\ref{cos-2}) (blue
    dot-dashed line).}
\end{figure}

The analysis of Eqs. (\ref{f}) together with the available numerical
solutions for $r=0.26$ and $3.86$ {\cite{Leal1971}} allowed us to
suggest the following approximation
\begin{equation}
  \label{approx-f}{ { f(C,r) \simeq  \frac{C/\!\left( r + 1/r \right)^2 }{\pi \left[ 1 + 4\,  C^2/\!\left( r + 1/r \right)^2  \right] ^{3/2}} }}\ .
\end{equation}
A comparison of this approximation with the exact numerical solution
provided in Ref. \cite{Leal1971} is presented in Fig.~\ref{fig:PDFs},
upper panel. One sees that Eq.~(\ref{approx-f}) fits the numerical
data with an accuracy better than $5\%$. This is more then enough for
our purpose to offer an approximate formula for $\langle \cos^2
\theta_{\rm N}\rangle$ as a function of $r$, see below.

Next we use the fact that Jeffery orbits do not intersect on the unit
sphere. In other words, fixing $C$ results in a relation between
$\theta$ and $\phi$ along an orbit. This relationship is obtained by
inverting Eq. (\ref{JeffOrbitsEqsTheta}):
\begin{equation}
  \label{conA-1}
  C= C(  \theta,   \phi)= \tan   \theta \sqrt{r^{-2}\sin^2   \phi + \cos^2   \phi  }\ .
\end{equation}
Substituting this function into any one of Eqs.~(\ref{f}) we get
$f(C,r)$ for a given regime of $r$. This is then substituted in
Eqs.~(\ref{PDF1}), leading finally to a solution of the orientational
PDF $\mathcal P( \theta, \phi)$, which can be used for averaging
Eqs.~(\ref{spheroid}) in the case of weak Brownian motions.  As a
consistency check of the approach one can consider the trivial case
$r=1$.  From Eq. (\ref{conA-1}) one finds $C=\tan \theta$, then from
Rq. (\ref{r=1}) $f(C)= \sin \theta \cos^2 \theta / 4 \pi $ and, finally
from (\ref{FP-2a}) $\mathcal P_C = 1/ \sin \theta \cos^2 \theta$. As
the result one has for sphere $\mathcal P=1/4 \pi$, as expected.

For moderate and strong Brownian rotational motion the notion of
separate Jeffery orbits becomes irrelevant. In this case we need to
solve Eq.~\eqref{FP} without approximations. Once this equation is
solved we can compute $\langle \cos^2 \theta\rangle$ and substitute
the answer in Eqs.~\eqref{spheroid}. To achieve this in the most
general case is not a simple task, and here we satisfy ourselves with
the two limiting cases of very large and very small rotational
diffusion. The second case was discussed above. For the case of very
strong rotational diffusion we can use the same Eqs.~(\ref{spheroid}),
but averaged with a uniform PDF $\mathcal P( \theta, \phi)=1/4
\pi$. This is because the very strong rotational diffusion
  tends to distribute particles motions around the Jeffery orbits
  uniformly. The corrections up to $\mathcal{O}(S/D_\mathrm{r})^2$ to
  such a uniform distribution may be found in Ref.~\cite{ML79,
    McMillen77}

\subsection{\label{ss:spheroids-R} Predictions of the model }

\subsubsection{\label{sss:SpheroidLargeDiff} {Spheroidal particle in a
    strong Brownian diffusion limit}}

With very strong Brownian diffusion, the particles are oriented
completely randomly, and $\langle \cos^2\theta_{\mathrm{N}}\rangle=
1/3$. In this case Eqs.~\eqref{K} and (\ref{spheroid}) give the
results reported in Fig.~\ref{f:7}, upper panels. The upper left panel
shows Nu vs. $r$ dependence for various $k$ from $1$ to $8600$ and
$\infty$ with volume fraction $\varphi=1\%$.  These results are rather
obvious: for $k=1$ one obtains Nu$=1$, i.e. no enhancement; the larger
the $k$, the larger the heat flux enhancement; for any finite $k$
there is a saturation of Nu$(r)$ for $r\to \infty$. {The value of
  $($Nu$-1)$ may be huge for spheroids (essentially, rod-like
  particles at $r\gg 1)$, e.g. for $k=8600$ { (diamond in water)},
  Nu$-1\simeq 30$, while for spherical particles $(r=1)$, Nu$-1=0.03$
  at the same volume fraction $\varphi=0.01$}. The values of Nu$(k,r)$
are bounded, i.e. $\mathrm{Nu}(1,r) \le \mathrm{Nu}(k,r) \le
\mathrm{Nu}(\infty,r)$.

The upper right panel shows Nu vs. $k$ dependence at $\varphi=1\%$ for
three values of $r$: $1$, $5$ and $10$. The values of Nu$(k,r)$ are
bounded, too: $\mathrm{Nu}(k,1) \le \mathrm{Nu}(k,r) \le
\mathrm{Nu}(r,\infty)$. Here, the more elongated particle is (larger
$r$), the better the enhancement.

And last, but not least: elongated particles may touch each other much
easier. As we show in the Appendix~\ref{aa:rnphi}, the basic
geometrical requirement that the mean inter-particle distance is less
than the largest particle size means that $r < r_0\equiv
\varphi^{-1/2}$, which is the basic criteria of the dilute limit. The
Nu$(r)$ dependence for $r < r_0$ is shown by solid curves, while the
region $r>r_0$ is shown by dashed curves in Figs.~\ref{f:7}, left
panels. Provided, $r \le 10$ at $\varphi=1\%$, the enhancement may be
still considered as large but not huge: for $k=5$, the saturation
level of (Nu-1) is about $0.25$ while for spheres Nu$-1\approx 0.017$.

The conclusion in the case of very strong Brownian diffusion is that
the particles with larger $k$ and $r$ bring larger heat flux
enhancement in the laminar shear flow.

\subsubsection{\label{sss:SpheroidSmallDiff} {Spheroidal particle in a
    weak Brownian diffusion limit}}

\begin{figure*}[t]
  \begin{center}
  \includegraphics[width=8.5cm ]{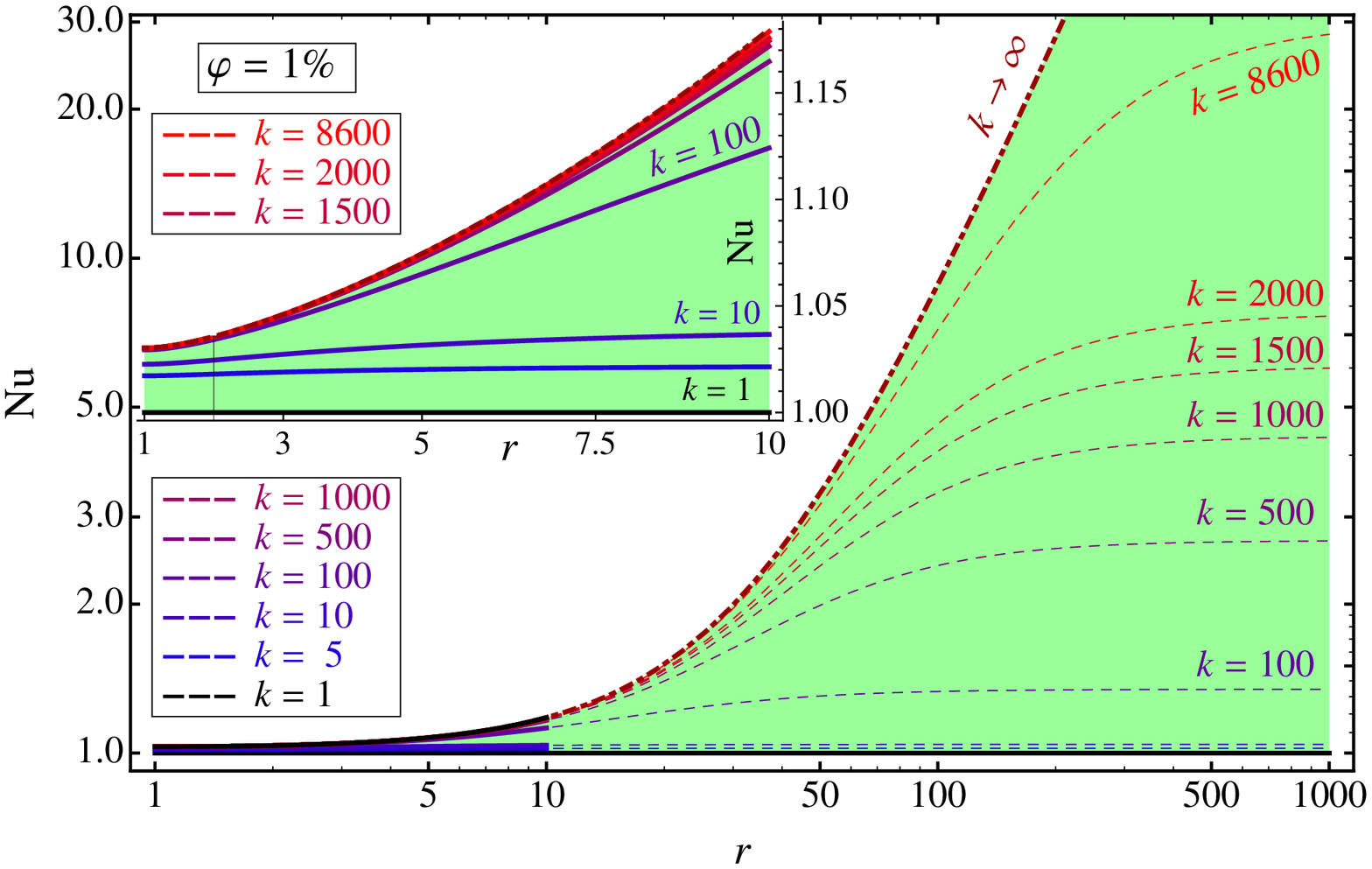}
  \includegraphics[width=8.5cm]{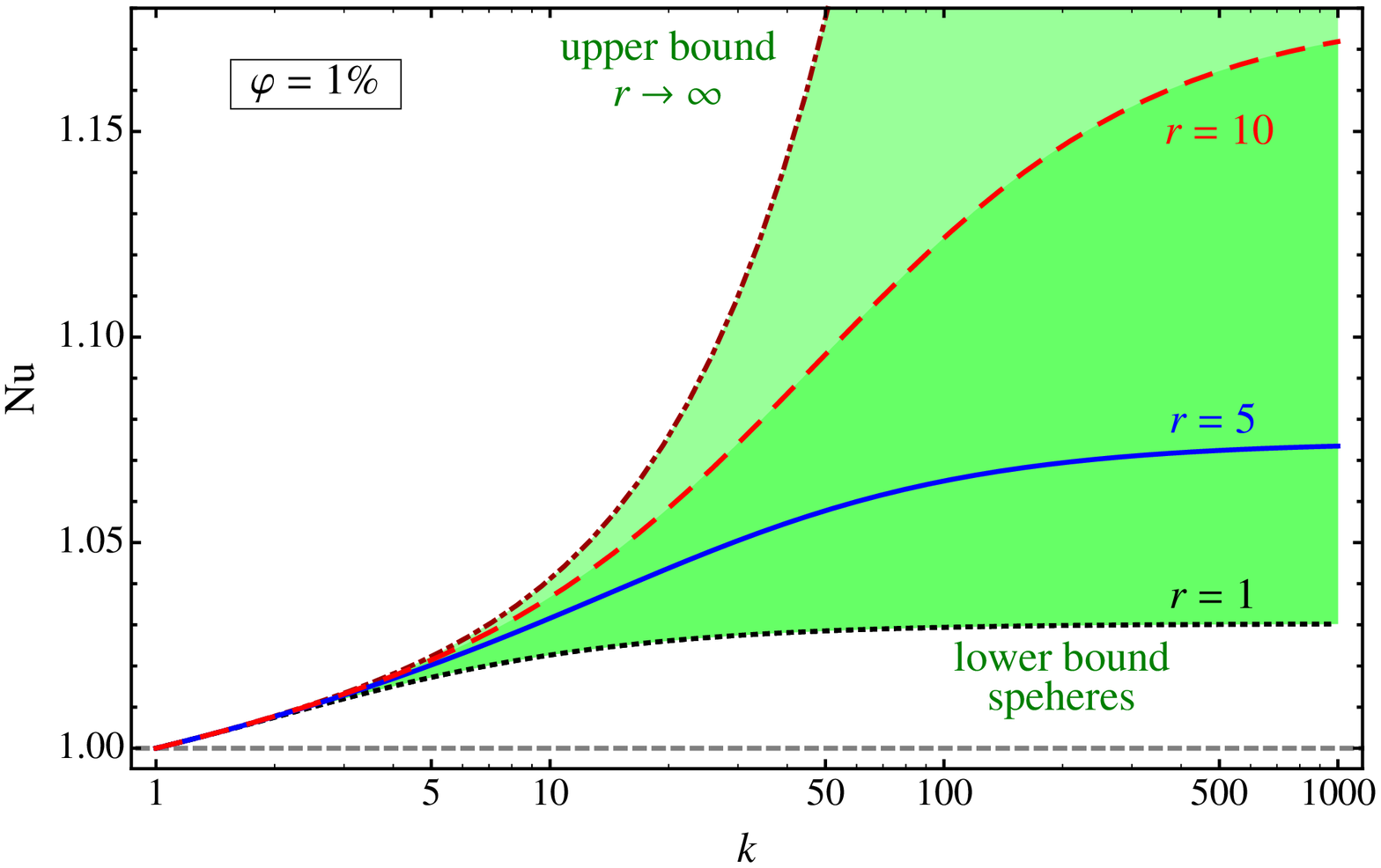}\\
  \includegraphics[width=8.5cm]{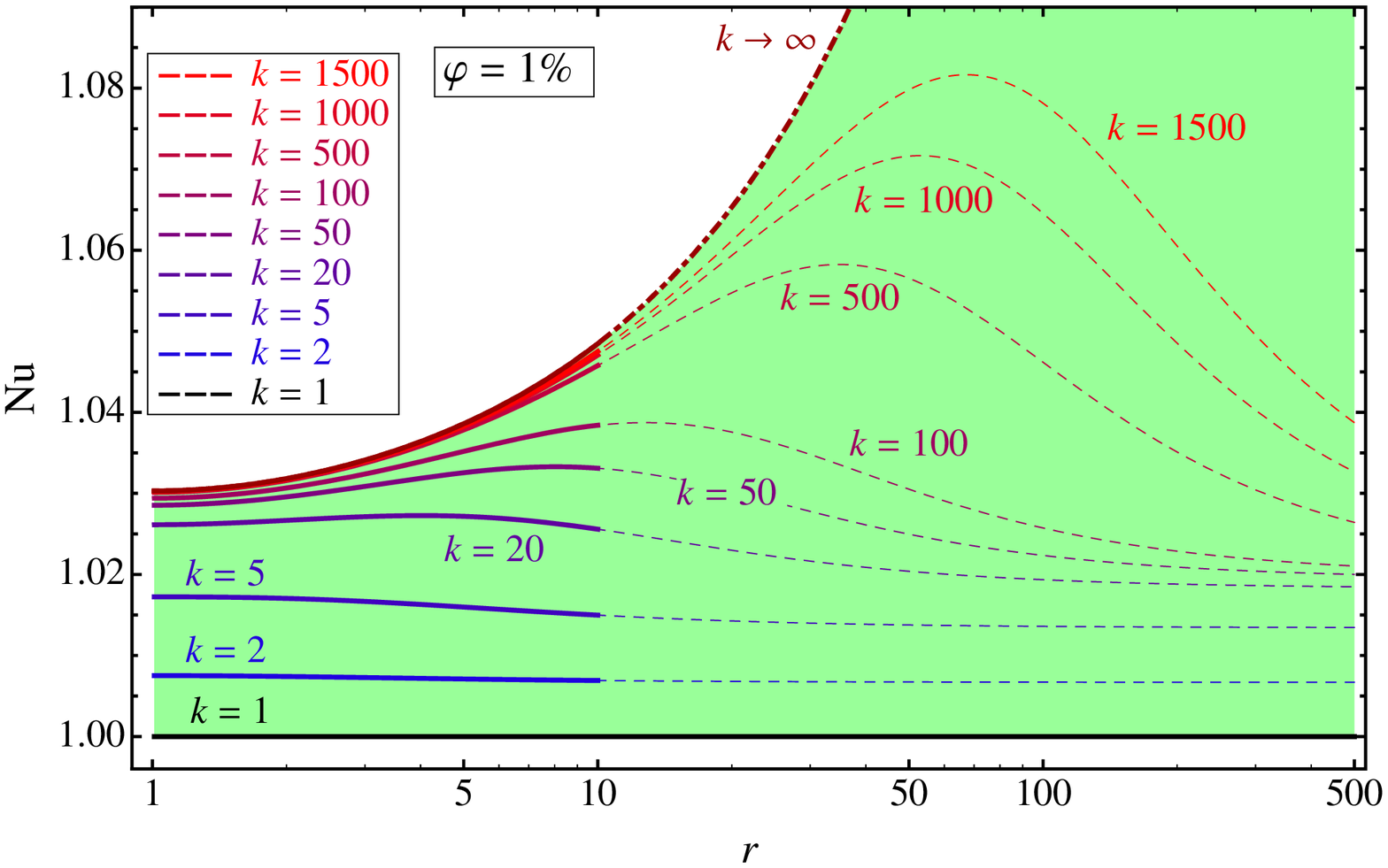}
  \includegraphics[width=8.5cm]{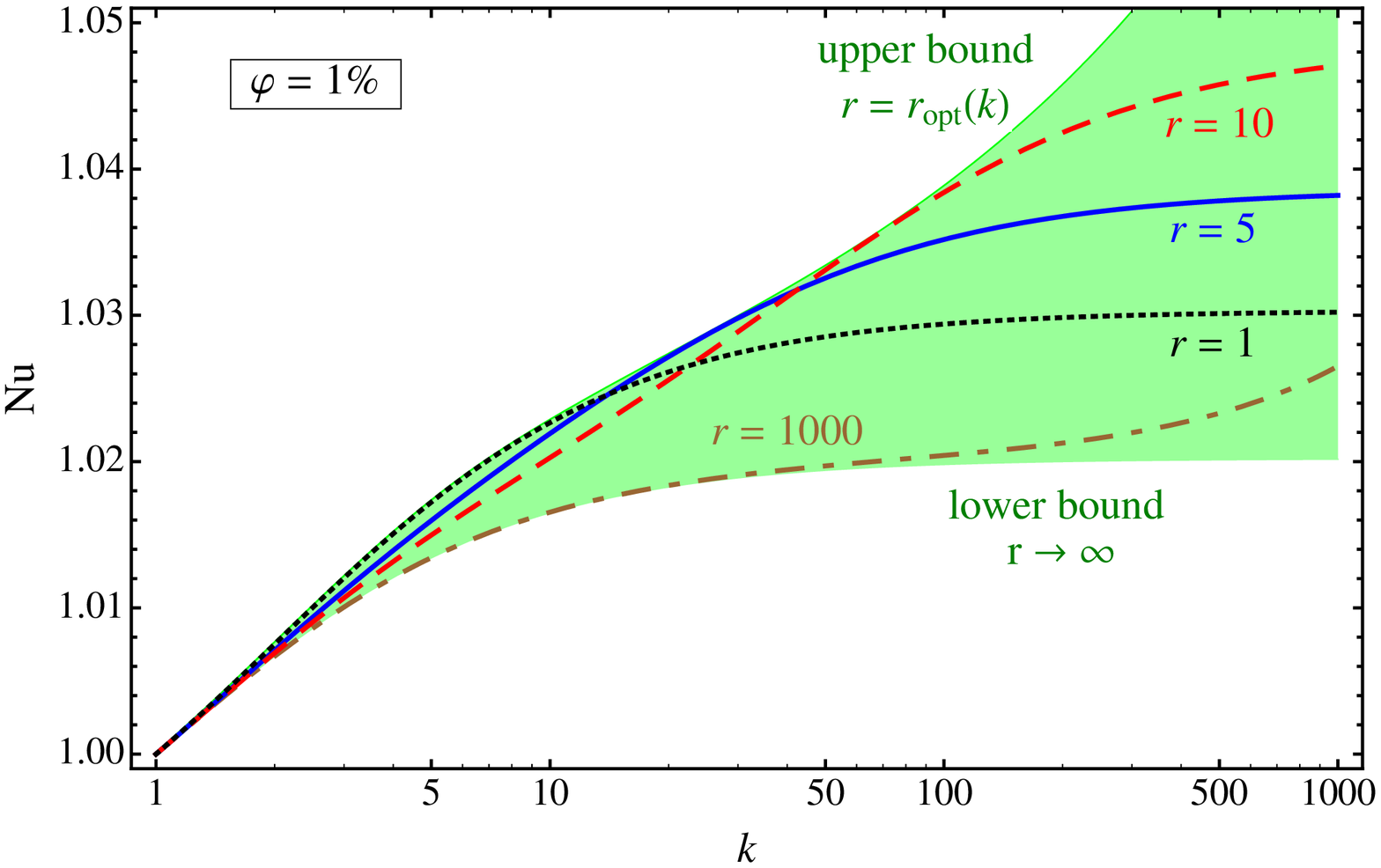}
  \end{center}
  \caption{\label{fig:Nu_Nan_Jeffery} (Color online). $\varphi = 1\%$,
    $r_0=10$.  \textbf{Upper panels:} \textit{Strong} Brownian
    diffusion limit: $\left\langle\cos^2\theta_\mathrm{N}
    \right\rangle\rightarrow 1/3$, \textbf{lower panels:}
    \textit{Weak} Brownian diffusion limit.  \textbf{Left panels:}
    Nusselt number, Nu, vs. particle aspect ratio, $r$, for various
    heat diffusivity ratios, $k$: solid curves for $r<r_0$, dashed
    curves for $r>r_0$ (the region, where the particles may, in
    principle, start touching each other). The shaded (green) area is
    the region of all possible Nu$(k,r)$ at $\varphi = 0.01$: the
    lower bound is given by Nu$(1,r)$, the upper one is
    Nu$(\infty,r)$.  \textbf{Right panels:} Nu vs. $k$ for various,
    $r$. Dotted (black) curve is for $r = 1$, spherical particles,
    solid (blue) curve -- $r = 5$, dashed (red) curve -- $r = 10$, and
    dot-dashed (brown) curve -- $r = 1000$, lower panel, or $r \to
    \infty$, upper panel. The shaded (green) area is the region of all
    possible Nu$(k,r)$ at $\varphi = 0.01$: the lower bound is
    Nu$[k,1]$ in upper panel and Nu$(k,\infty)$ in lower panel, while
    the upper bound is Nu$(k,\infty)$ in upper panel and
    Nu$[k,r_\mathrm{opt}(k)]$ in lower panel, where
    $r_\mathrm{opt}(k)$ maximizes Nu at every $k$, see more in the
    text. The geometrical constrain imposes $1\le r \le
    \max(r_\mathrm{opt}, 10)$ for $\varphi = 1\%$. \label{f:7} }
\end{figure*}

To complete the calculations of Nu$(r)$ dependence in a weak
{Brownian} diffusion limit in the framework of model~(\ref{spheroid}),
we have to find $\langle \cos^2 \theta_{\mathrm{N}} \rangle$
vs. $r$. To make a long story short, we compared in Fig.~\ref{fig:PDFs},
lower panel, the dependence of $\langle \cos^2
\theta_{\mathrm{N}}\rangle$ vs. $r$ obtained by numerically solved
PDF, Eq.~(\ref{FP-1a}) (see Refs.~\cite{Leal1971, Hinch1972} for more
details), shown by (black) solid curve, and by approximate analytical
PDF, Eq.~(\ref{approx-f}), shown by (red) dashed curve. As one sees
these two dependence coincide within line width. Moreover, by careful
analysis of various limiting cases, we suggested the following simple
model dependence
\begin{equation}
  \label{cos-2}
  { \langle \cos^2 \theta_{\mathrm{N}} \rangle\simeq \frac 2 {6+\pi\, (r-1)}\,,}
\end{equation}
shown in Fig.~{\ref{fig:PDFs}}, lower panel, as a (blue) dash-dotted
curve.  Eq.~(\ref{cos-2}) fits the exact dependence with an accuracy
of 5\%. Therefore, Eqs.~(\ref{spheroid}) and (\ref{cos-2}) can be used
in our analysis to make predictions on the thermal properties in the
limit of small Brownian diffusion of fluids laden with spheroidal
nano-particles of different aspect ratios and different thermal
conductivity ratios with Peclet number up to unity. Corresponding
results are shown in Fig.~\ref{fig:Nu_Nan_Jeffery}, lower panels.


The lower left panel in Fig.~\ref{fig:Nu_Nan_Jeffery} shows Nu$(r)$ for
different $k$ and $\varphi=0.01$, at which the Maxwell-Garnett limit
of Nu$-1$ for spherical particles ($r=1$) is $3\varphi=0.03$. Notice,
the Nu$(r)$ dependence is not monotonic and has a maximum at some
$r=r_\mathrm{opt}$, which depends on $k$. The reason for this is the
competition of two effects: more elongated nano-particles give larger
contribution to the heat flux {when their longer axis is aligned with}
the temperature gradient, {which is orthogonal to the velocity
  gradient (shear) in our case}. However longer nano-particles are
affected more readily by the shear, which tends to orient them in the
unfavorable direction orthogonal to the temperature gradient;  then
their contribution to the heat flux is even less than the one of the
spherical particles (cf. Fig.~\ref{NuVsTheta}).  Again, the values of
Nu$(k,r)$ are bounded, i.e. $\mathrm{Nu}(1,r) \le \mathrm{Nu}(k,r) \le
\mathrm{Nu}(\infty,r)$.

The lower right panel in Fig.~\ref{fig:Nu_Nan_Jeffery} shows Nu$(k)$ for
different aspect ratios, $r$ at $\varphi = 1\%$. This is again a
consequence of the above described competition. The values of Nu are
bounded by $\mathrm{Nu}(k,\infty) \le \mathrm{Nu}(k,r) \le
\mathrm{Nu}[k,r_\mathrm{opt}(k)]$. Moreover, for $1 \le k \lesssim 7$
the optimal nano-particle shape is spherical.

\begin{figure*}[t]
  \begin{center}
   \includegraphics[width=8.5cm]{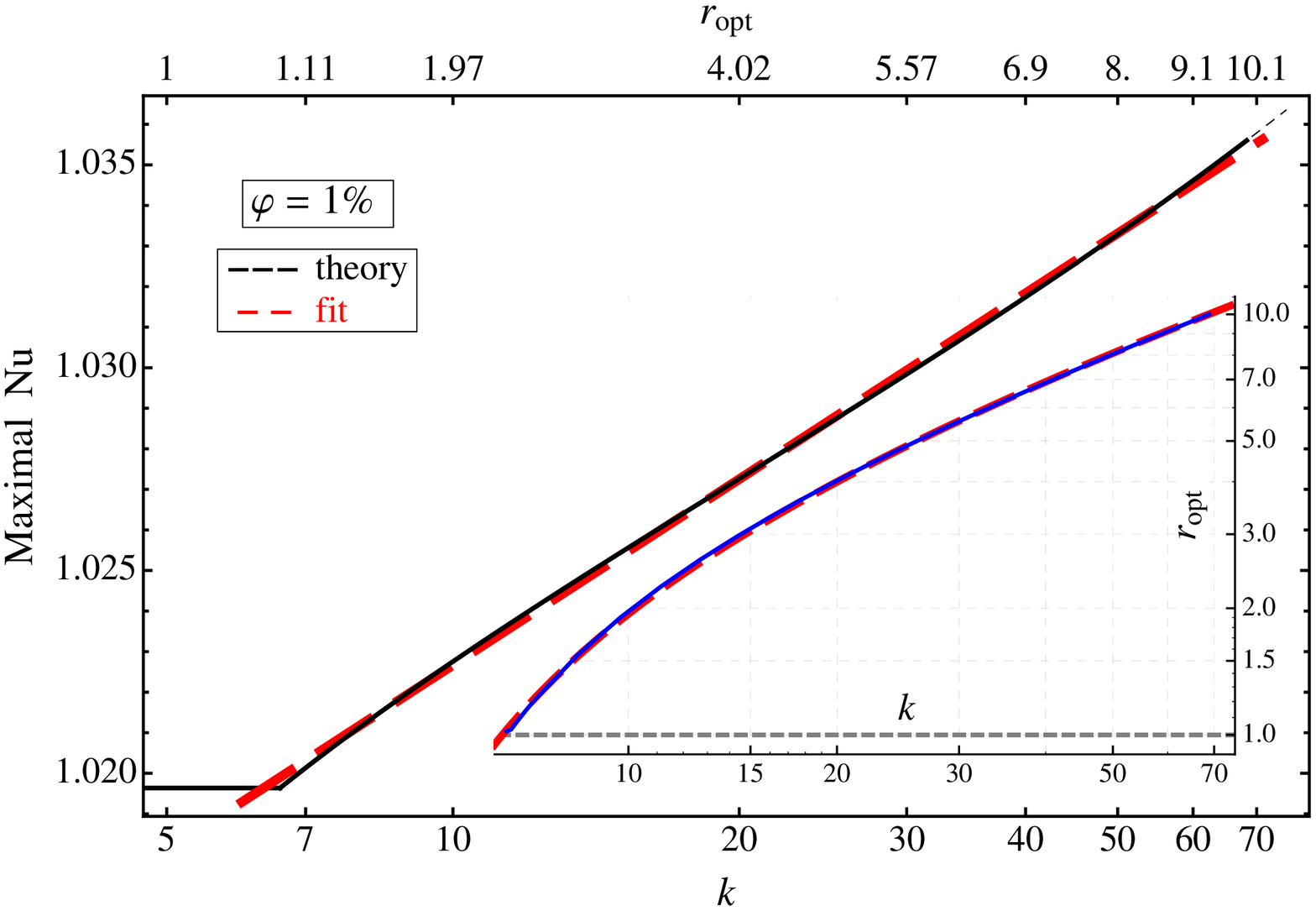}
   \includegraphics[width=8.5cm]{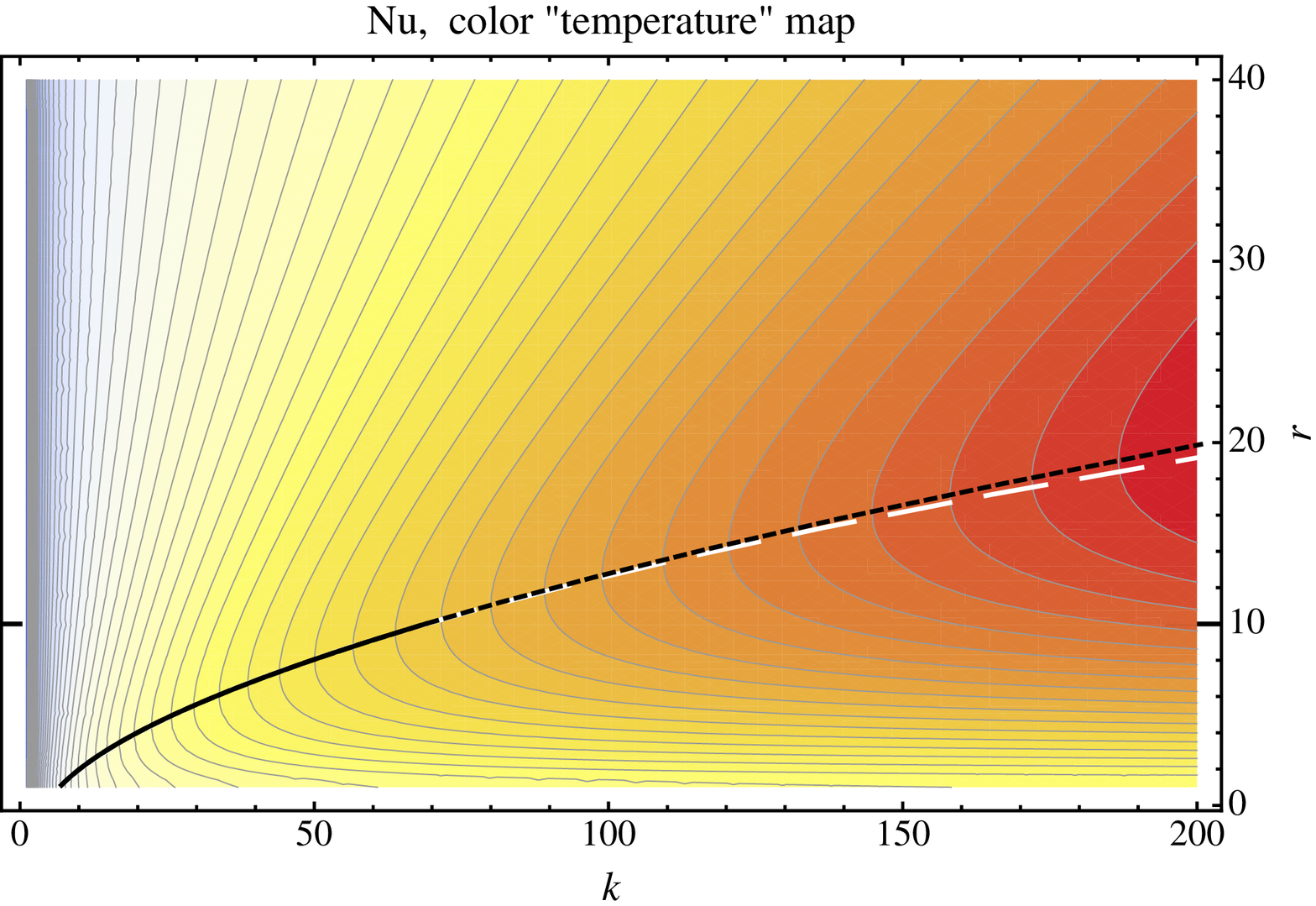}
  \end{center}
  \caption{\label{fig:OptimalNu} (Color online). \textit{Weak}
    Brownian diffusion limit, $\varphi = 1\%$, $r_0=10$.
    \textbf{Left:} The largest reachable Nusselt number,
    $\mathrm{Nu}(k, r_\mathrm{opt})$, solid (black) curve
    ($r_\mathrm{opt}$ maximizes Nu for a given $k$). Dashed (red) line
    is the fit by $ \mathrm{Nu}_\mathrm{max}^{\mathrm{fit}} \approx 1
    + \varphi [0.75 +0.66 \ln(k)]$. \textbf{Insert:} $r_\mathrm{opt}$
    \textit{vs}. $k$, solid (blue) curve, and the fit
    $r_\mathrm{opt}^{\mathrm{fit}} \approx -3.03 + 1.57 \sqrt{k}$ for
    $1 \le r_\mathrm{opt} \le 10$, dashed (red) line. Since $\varphi =
    0.01 \ll 1$, this fit may be considered as $\varphi$-independent
    for $\varphi \le 1\%$.  \textbf{Right:} Contour-density plot of
    Nu$(k,r)$ color-codded with the ``temperature'' map: the larger
    the value of Nu, the wormer the color; contours show iso-values of
    Nu. Thick (black) solid and dashed curves represent those pairs of
    $(k,r)$ for which Nu is maximal (i.e. solid (blue) curve in the
    Insert); solid curve is for $r < r_0$, and the dashed one is for
    $r > r_0$. Wide-dash (white) curve is the
    $r_\mathrm{opt}^{\mathrm{fit}}(k)$ fit.  }
\end{figure*}

As seen in Fig.~\ref{fig:Nu_Nan_Jeffery}, lower left panel, there
exists a maximum of Nu$(r=r_\mathrm{opt})$ for a given $k$. This
maximal Nusselt number at its maximizing (optimal)
$r=r_\mathrm{opt}(k)$ is shown in Fig.~\ref{fig:OptimalNu}, left, as a
function of $k$ for $\varphi = 0.01$. For $k > 7$, the maximal Nu
behaves like
\begin{subequations}
  \begin{eqnarray}
    \label{MaxNuFit}
    \mathrm{Nu}_\mathrm{max}^{\mathrm{fit}} \approx 1 + \varphi \left[0.75 +0.66 \ln(k) \right], \ \ \varphi = 0.01\ .
\end{eqnarray}
Since here $\varphi \ll 1$, the part in parenthesis may be considered
as $\varphi$-independent, thus, the fit~(\ref{MaxNuFit}) may be used
at any $\varphi \le 1\%$.

The right panel of Fig.~\ref{fig:OptimalNu} exhibits a contour-density plot
of Nu$(r,k)$ at $\varphi = 0.01$. Thick (black) solid and dashed
curves show $r_{\mathrm{opt}}(k)$ dependence (solid curve is for $r <
r_0$, and the dashed one is for $r > r_0=\varphi^{-1/2} = 10$). The
inverse dependence $k(r_{\mathrm{opt}})$ is easy to obtain in a
symbolic computation software by solving
$\partial$Nu$(r,k,\varphi)/\partial r
= \partial\mathcal{K}(r,k,\varphi)/\partial r = 0$ at $r =
r_{\mathrm{opt}}$, though, the answer appears to be very cumbersome to
be shown here.  However, our analysis reveals that $r_{\mathrm{opt}}
\sim \sqrt{k}$, and for $\varphi =0.01$ and $1 \le r_{\mathrm{opt}}
\le 10$, .
\begin{eqnarray}
  \label{RoptFit}
  r_\mathrm{opt}^{\mathrm{fit}}(k) \approx -3.03 + 1.57 \sqrt{k}\,, \quad \varphi = 0.01\ .
\end{eqnarray}
\end{subequations}
This dependence is shown in Fig.~\ref{fig:OptimalNu}, right, as a
wide-dashed (white) curve, which deviates from the analytical solution
$r_\mathrm{opt}(k)$ for $r_{\mathrm{opt}} > 10$. Again, since $\varphi
\ll 1$, the fit~(\ref{RoptFit}) may be used\footnote{According to
  Eqs.~(\ref{NuSpherois}) and (\ref{spheroid}) for $\varphi \ll 1$,
  $\partial\mathrm{Nu}/\partial r \simeq \partial \left[ \beta_1
    +(\beta_2 -\beta_1)\langle \cos^2 \! \theta_{\rm N} \rangle
  \right] /\partial r$, which is a function of $(k,r)$ only.} at any
$\varphi \le 1\%$.

In conclusion, we should notice that the rich information about heat
flux enhancement, shown in the four panels of
Fig.~\ref{fig:Nu_Nan_Jeffery}, is just an illustration of the analytical
dependence Nu$(\varphi, k, r)$ given by the analytical
expressions~(\ref{spheroid}) and (\ref{cos-2}).  This is an important
result of our modeling.

\section*{CONCLUSIONS}

We presented a study of the physics of the heat flux in a
fluid laden with nanoparticles of different physical properties
(shape, thermal conductivity, etc). We developed a new analytical
model for the effective thermal properties of dilute nanofluid
suspensions. Our model accounts for nanoparticle rotation
dynamics including the fluid motion around the nanoparticles.  We note that our model reproduces the
classical Maxwell-Garnet model in the appropriate static limits.

We used a combination of theoretical models and numerical experiments
in order to make progress from the simplest case of spherical nanoparticles
in a quiescent fluid to the most general case of rotating spheroidal
particles in shear flows. The new physical ingredient that we consider
is the exact dynamics of particles in shear flows. This constitute a
novelty as most of the models introduced so far, to explain the
thermal properties of thermal colloids, have focused only on the
static properties of the nanoparticle suspension.  Our model starts
from the realization that particles (spherical or spheroidal) in
the presence of a gradient of the velocity field are induced to
rotate. The dynamics of rotation is absolutely non trivial,  but it has
been studied at length with correspondece (for the case of a laminar and
stationary shear flow) to the Jeffery orbits.

The particles rotation dynamics has a double influence on the thermal
properties of the nanofluid. First, particles rotation induces fluid
motions in the proximity of the particles, this in turn can enhance
the thermal fluxes by means of advective motions along the direction
of the temperature gradient. Second, the Jeffery dynamics of
particles leads to a statistical distribution of particles orientation
that depends on a multitude of parameters, e.g. the particle aspect
ratio, the shear intensity as well as on the intensity of thermal
fluctuations. The statistical distribution of particle orientation
has a dramatic influence on the heat flux: an elongated particle
oriented along the temperature gradient increases the thermal flux,
while a particle with perpendicular orientation reduces it.

The statistical orientation of particles can thus produce a mixed
effects with a non-trivial dependence on the particle aspect
ratio. More Elongated particles can enhance the heat flux because of
the stronger contribution when properly aligned to the temperature
gradient but, because of shear, more elongated particles are also
spending more time in the unfavorable direction (i.e. perpendicular to
the temperature flux) thus reducing the thermal conductivity of the fluid.

By means of numerical approximations we are able to provide closed
expressions for the effective conductivity of the fluid under several
flow regimes and for several physical parameters. Our model
considerably extends classical models for nanofluid heat transfer,
like e.g. the one of Maxwell-Garnet, and may help to rationalize some
of the recent experimental findings. In particular, we suggest that
experiments should consider more carefully  measurements performed
in quiescent and under flowing conditions: the particles dynamics
may lead to very different thermal properties in the two cases.

Finally, the next steps toward a robust predictive models for the heat
transfer in nanofluids should include the effect of surface (Kapitza)
resistance and the effect of nanoparticle aggregation.  Further it
would also be extremely important to extend the model to the case of
heat flux in turbulent nanofluids as this case is very relevant to
many applications.  In the presence of turbulence a particular attention
should also be paid to the effect on the drag induced by the presence
of spherical, rod-like or maybe even deformable nanoparticle
inclusions.

%
%

\section*{Acknowledgments}
We acknowledge financial support from the EU FP7 project ``Enhanced
nano-fluid heat exchange'' (HENIX) contract number 228882.

\appendix

\section{\label{ss:Numerics}Numerical approach}

The numerical simulation of the conjugated heat transfer problem,
equations (\ref{Eq1}-\ref{Eq3}), is performed by means of two coupled
D3Q19 Lattice Boltzmann (LB) equations under the BGK approximation
\cite{Succi2001} (for velocity and temperature fields) and
Molecular-Dynamics simulations (for particles motion):
\begin{subequations}
\label{LBEQ}
\begin{eqnarray}
  \label{LBEQ1}
  f_{l}({\bm x}+{\bm c}_l,t+1)-f_{l}({\bm x},t)=&\dfrac{ f_l^{\rm eq}({\bm x},t) -f_l({\bm x},t) }{\tau_{\rm f}}\,,~~~~  \\
  \label{LBEQ2}		
  g_{l}({\bm x}+{\bm c}_l,t+1)-g_{l}({\bm x},t)=&\dfrac{ g_l^{\rm eq}({\bm x},t) -g_l({\bm x},t) }{\tau_{\rm g}}\,,~~~~
\end{eqnarray}
\end{subequations}
where $f_{l}({\bm x},t)$ is the Lattice Boltzmann distribution
function for particles at $({\bm x},t)$ with velocity ${\bm c}_l$
(with $l={0,\dots,18}$ for D3Q19), and $f_{l}^{\rm eq}$ is its
equilibrium distribution; $g_{l}({\bm x},t)$ is the distribution
functions associated with the temperature and $g_{l}^{\rm eq}$ is its
equilibrium distribution.  The first LB, Eq.~(\ref{LBEQ1}), evolves
the fluid flow outside of the rigid particle and its momentum is
coupled with the particle boundaries by means of a standard scheme, as
proposed by Ladd \cite{Ladd2006}. The second LB, Eq.~(\ref{LBEQ2}),
evolves the temperature field, treated as a passive scalar as proposed
in \cite{Chen1998}, solving the conjugated heat transfer problem
simply by means of adjusting the thermal conductivity to the correct
values in the fluid and inside the particle [eqn. (\ref{Eq1}) and
(\ref{Eq2})]. Thermal and velocity boundary conditions, at the top and
bottom walls, impose the LB populations to equal the equilibrium
populations (corresponding to the desired velocity and
temperature). This approach can produce small temperature and velocity
slip which are kept into account by measuring the effective
temperature and velocity profiles, thus increasing the accuracy. The
code employed is fully parallelized by means of MPI libraries
\cite{MPI} thus allowing large system sizes, important to study the
influence of finite size effects.

Density, momentum and temperature are defined locally at $({\bm x},t)$
as coarse-grained (in velocity space) fields of the distribution
functions
\begin{eqnarray}
  \label{HYDRO}
  \rho_{\rm f}=\sum_{l}f_{l}\,, \hspace{.2in} \rho_{\rm f} {\bm u_{\rm f}}=\sum_{l}{\bm c}_l f_{l}\,, \hspace{.2in}  \rho_{\rm f} T_{\rm f}=\sum_{l} g_{l}\ .~~~~~
\end{eqnarray}
A Chapman-Enskog expansion \cite{Guo2} around the local equilibria
$f^{\rm eq}_l({\bm u},\rho)$ and $g^{\rm eq}_l({\bm u}, T)$
\cite{Guo1} leads to the equations for temperature and momentum
(\ref{Eq1})-(\ref{Eq2}): the streaming step on the left hand side of
(\ref{LBEQ1}) reproduces the inertial terms in the hydrodynamical
equations, whereas the diffusive terms (dissipation and thermal
diffusion) are closely connected to the relaxation (towards
equilibrium) properties in the right hand side, with $\nu$ and $\chi$
related to the relaxation times $\tau_{\rm f}$, $\tau_{\rm g}$
\cite{Succi2001}.

\subsubsection{\label{ss:LnL}Conjugated heat transfer for a spherical particle at
  rest}
Consider a spherical particle of radius $R$ immersed in a quiescent
fluid, in which a constant temperature gradient, $\bm  G$, is
maintained.  The temperature distribution is \cite{LnL}:
 \begin{eqnarray}
   \label{eq:LnL}
  T_{\rm{p}}(\bm  r) &=& \frac{3}{2+k}\bm  G \cdot \bm  \rho\,, \qquad\qquad\qquad\ \ \, r < R\,, \\
  T_{\rm{f}}(\bm  r) &=& \left[1 + \frac{ 1-k }{2+k}\left( \frac{R}{\rho}\right)^{\!3} \right]\bm  G \cdot \bm  \rho\,, \quad r \geq R\,, \nonumber
\end{eqnarray}
where $\rho = \sqrt{x^2+y^2+z^2}$ is the distance from the particle's center.

In our Couette flow simulations, $\bm G = {\widehat{\bf{y}}}\,
\Delta/H$, but the temperature boundary conditions are different from
\cite{LnL}. One should expect then deviations close to the walls
especially for large particles. The results are presented in
Fig.~\ref{f:2}.

\begin{figure}[t]
  \begin{center}
   \includegraphics[width=8.5cm]{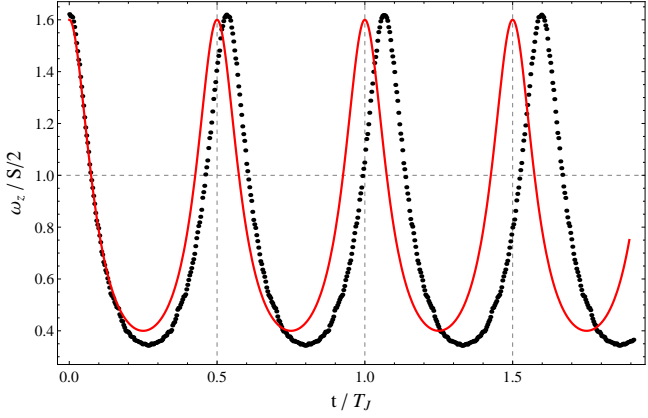}
  \end{center}
  \caption{\label{Jeffery_Test} {Angular velocity \textit{vs.}
      time. Solid (red) curve -- theory of Jeffery \cite{Jef},
      Eqs.~(\ref{eqs:JefferyTheory}), for an infinite domain with a
      constant simple shear at infinity and the {creeping} (Re$
      \rightarrow 0$) flow around the particle, dots -- numerical
      results for small but finite Re$ = 0.05$. Configuration: $64^3$,
      Pr~$ =10$, $k = 1$, $r=2$ ($a=14$, $b=7$), Pe~$ =0.5$.}}
\end{figure}

\subsubsection{Spheroid in a shear flow -- ``Jeffery'' test}
A spheroidal particle of aspect ratio $r$ in a simple shear flow
undergoes a spinning motion (in the $XY$-plane). The angular velocity
and the period of such a motion were predicted by Jeffery \cite{Jef}
for a case of a creeping flow around the particle, i.e. when
Re$\rightarrow 0$:
\begin{subequations}
\label{eqs:JefferyTheory}
\begin{eqnarray}
  \label{eq;JefferyAngularVel}
  \omega_z &=& \frac{S\,r /\! \left(r+1/r\right)}{\cos^2\!\left(2\pi\, t/T_{\rm J} \right) + r^2 \sin^2\!\left(2\pi\, t/T_{\rm J} \right)}\,,~~~~ \\
  \label{eq;JefferyPeriod}
  T_{\rm J} &=& {2\pi}\left(r+{1}/{r}\right)/{S}\ .
\end{eqnarray}
\end{subequations}
For non-vanishing Re, the actual period of rotation of the particle
deviates from that predicted by Jeffery. In Fig.~\ref{Jeffery_Test} we
compare the results of our LBM simulations with
Eq.~(\ref{eq;JefferyAngularVel}) for the case of $r=2$ and Re$ =
0.05$. There is a 6\% difference in periods due to a finite Re, but
also due to influence of the other particles present due to the
periodic b.c., and also the influence of the walls ($2a/H \simeq
0.44$).

\section{\label{aa:rnphi} Approximation of non-interacting particles}
The approximation of non-interacting spheroids is comming from simple
geometrical considerations, and it is valid provided the particle
aspect ratio $r < \varphi^{-1/2}$, as confirmed by the following
derivation:
\begin{eqnarray}
	\varphi &=& \frac{V_{\rm {\rm all\,paricles}}}{V_{\rm total}} = \frac{N\,V_{\rm particle}}{V_{\rm total}} = \frac{V_{\rm particle}}{V_{\rm eff}} \nonumber\\
	&=& \frac{4\pi\,a\,b^2/3}{V_{\rm eff}}  = \frac{4\pi\,a\,b^2/3}{\mathcal{L}^3}   = \frac{4\pi\,a^3\,r^{-2}/3}{\mathcal{L}^3}\quad \Rightarrow \nonumber\\
	\mathcal{L} &=& \sqrt[3]{\frac{4\pi\,a^3\,r^{-2}/3}{\varphi}}  = a\sqrt[3]{\frac{4\pi}{3\,r^2 \varphi}}\,, \nonumber \\
	2a &<& \mathcal{L} \quad \Rightarrow \quad r < \sqrt{\frac{\pi}{3 \varphi}} \quad \Rightarrow \quad r < \varphi^{-1/2}.
\end{eqnarray}
Here $N$ is the number of particles in the total volume (of fluid and
particles together), $V_{\mathrm{total}}$, $V_{\mathrm{eff}} =
V_{\mathrm{total}}/N$ is an ``effective'' volume per particle, and
$\mathcal{L} = V_{\mathrm{eff}}^{1/3}$ is a characteristic
length/dimension of the effective box/volume embedding the particle.

\end{document}